\documentclass[final,3p,times]{elsarticle}

\usepackage{graphicx}
\usepackage{newtxtext}
\usepackage{newtxmath}
\usepackage{amsmath}
\usepackage{natbib}
\usepackage{hyperref}
\usepackage{graphicx}
\usepackage{subcaption}
\usepackage[utf8]{inputenc}
\hypersetup{
    colorlinks = true,
    urlcolor   = blue,
    citecolor  = black,
}

\newcommand{\RomanNumeralCaps}[1]
\linenumbers

\usepackage[T1]{fontenc}
\usepackage{float}
\usepackage{acronym}

\begin{document}
\acrodef{LBM}[LBM]{lattice Boltzmann method}
\acrodef{NN}[NN]{neural network}
\acrodef{CBC}[CBC]{characteristic boundary condition}
\acrodef{CFD}[CFD]{computational fluid dynamics}
\acrodef{BGK}[BGK]{Bhatnagar-Gross-Krook}
\acrodef{MRT}[MRT]{multi-relaxation-time}

\begin{frontmatter}

\title{Reduction of Outflow Boundary Influence on Aerodynamic Performance using Neural Networks}

\author[uq,siegen,hbrs]{Mario C. Bedrunka\texorpdfstring{\corref{cor1}}{*}} 
\author[hbrs,fu]{Dirk Reith} 
\author[siegen]{Holger Foysi}
\author[warsaw]{\L{}ukasz \L{}aniewski-Wo\l{}\l{}k}
\author[uq]{Travis Mitchell}
\cortext[cor1]{Corresponding author. E-mail address: bedrunka.research@gmail.com}

\affiliation[uq]{organization={School of Mechanical and Mining Engineering, The University of Queensland},
            addressline={}, 
            city={St Lucia},
            postcode={4072}, 
            state={QLD},
            country={Australia}}
\affiliation[siegen]{organization={Chair of Fluid Mechanics, University of Siegen},
            addressline={Paul-Bonatz-Straße 9-11}, 
            city={Siegen-Weidenau},
            postcode={57076}, 
            state={},
            country={Germany}}
\affiliation[hbrs]{organization={Institute of Technology, Resource and Energy-efficient Engineering (TREE), \\ Bonn-Rhein-Sieg University of Applied Sciences},
            addressline={Grantham-Allee 20},
            city={Sankt Augustin},
            postcode={53757},
            state={},
            country={Germany}}
\affiliation[fu]{organization={Fraunhofer Institute for Algorithms and Scientific Computing (SCAI)},
            addressline={Schloss Birlinghoven},
            city={Sankt Augustin},
            postcode={53754},
            state={},
            country={Germany}}
\affiliation[warsaw]{organization={Institute of Aeronautics and Applied Mechanics, Warsaw University of Technology},
            addressline={Nowowiejska 24},
            city={Warszawa},
            postcode={00-665},
            state={},
            country={Poland}}

\begin{abstract}
The accurate treatment of outflow boundary conditions remains a critical challenge in computational fluid dynamics when predicting aerodynamic forces and/or acoustic emissions. This is particularly evident when employing the \ac{LBM} as the numerical solution technique, which often suffers from inaccuracies induced by artificial reflections from outflow boundaries. This paper investigates the use of \acp{NN} to mitigate these adverse boundary effects and enable truncated domain requirements. Two distinct \ac{NN}-based approaches are proposed: (1) direct reconstruction of unknown particle distribution functions at the outflow boundary; and (2) enhancement of established \acp{CBC} by dynamically tuning their parameters. The direct reconstruction model was trained on data generated from a 2D flow over a cylindrical obstruction. The drag, lift, and Strouhal number were used to test the new boundary condition. We analyzed results for various Reynolds numbers and restricted domain sizes where it demonstrated significantly improved predictions when compared with the traditional Zou \& He boundary condition. To examine the robustness of the \ac{NN}-based reconstruction, the same condition was applied to the simulation of a NACA0012 airfoil, again providing accurate aerodynamic performance predictions. The neural-enhanced \ac{CBC} were evaluated on a 2D convected vortex benchmark and showed superior performance in minimizing density errors compared to \acp{CBC} with fixed parameters. These findings highlight the potential of \ac{NN}-integrated boundary conditions to improve accuracy and reduce computational expense of aerodynamic and acoustic emissions simulations with the \ac{LBM}.
\end{abstract}
\acresetall



\end{frontmatter}

\section{Introduction} \label{sec:introduction}
Accurate prediction of aerodynamic forces and acoustic emissions in \ac{CFD} requires solving the flow field with both physical fidelity and numerical stability. Traditionally, the simulation of flow around objects, such as airfoils or bluff bodies, has been a cornerstone of \ac{CFD}, enabling the evaluation of lift and drag coefficients without resorting to physical prototypes \citep{zhang2024_1}. However, capturing not only the mean force coefficients, but also transient flow features such as pressure waves and vortex shedding becomes critical in applications involving noise prediction, unstable aerodynamics, and/or fluid-structure interaction. One persistent challenge for these simulations is the treatment of domain boundaries. If boundary conditions are not sufficiently nonreflecting, artificial waves can be introduced or reflected back into the domain, corrupting both aerodynamic and acoustic observables \citep{prosser2005}. This issue becomes even more pronounced when attempting to resolve acoustic fields, where spurious reflections can severely distort wave propagation and pressure fluctuation patterns. 

The accurate definition of boundary conditions is essential in high-fidelity simulations of flow phenomena, especially if physical effects or complexity occur close to the prescribed condition. This is evident in many natural and industrially relevant cases such as a turbulent vortex wake propagating through a simulation boundary, or acoustic waves interacting with it. This is exacerbated in simulations of acoustic fields where reflections are unwanted and need to be minimized for accurate predictions.
For the more conventional finite difference or finite volume approaches to resolving the Navier-Stokes equations, characteristic or non-reflecting boundary conditions have been previously developed and successfully applied. This came after the realization that even when extrapolated supersonic boundary conditions were used, numerical waves are able to backpropagate as they are reflected at, for example, regions of grid stretching or when one-sided differences are used~\citep{Vichne:1981,Vichne:1987}.
One of the first studies to consider non-reflecting boundary conditions for hyperbolic conservation laws was by Thompson~\cite{Thompson:1987}. Based on this work, Poinsot and Lele~\cite{Poinsot:1992} tried to extend the knowledge to the Navier-Stokes equations to establish well-posedness. This was done using the foundational work of Strikwerda~\cite{Strikwerda:1977} to specify the required number of boundary conditions. The general procedure consisted of casting the Navier-Stokes equations in a characteristic form, making use of a diagonalization of the flux Jacobians in using its eigenvectors. As only one Jacobian can be diagonalized at a time, a different set of equations must be derived for each boundary direction.
The characteristic form includes the time-varying wave amplitudes involving the eigenvalues which allows components entering the domain to be identified and modified appropriately. Based on the seminal works mentioned above, many extensions have been developed, including those for reacting flows, streamlines or using adjoint boundary condition formulations for optimization~\citep{Baum:1994,Sesterhenn:2001,Sutherland:2003,Colonius:2004b,WangM:2006,Yoo:2007,Lodato2008,Albin2011,Marinc2012}.

Other approaches make use of an asymptotic form of the equations based on a fictive center of propagation for the acoustic far-field~\citep{Tam:1998}, or use sponge layers to avoid reflections~\citep{Freund:1997,BodonySponge:2005,Mani:2012}. The combinations of sponge layers and characteristic boundary conditions can be commonly found in various fluid flow applications~\citep{Schaupp:2012,Foysi:2010a}. However, optimizing boundary condition treatment to further improve simulation fidelity and efficiency is still a significant research focus within the \ac{LBM}, as reflected in current literature~\citep{zhang2024_2, daeian2025, wang2025}. For the \ac{LBM}, \acp{CBC} have been sparsely applied, however, not to the extent found in conventional formulations of the governing equations. Seminal works include the analysis by Izquierdo and Fueyo~\cite{LBMCBC:2008} which used a velocity bounce-back boundary condition at the inlet and a characteristic-based boundary condition at the outlet. The characteristic directions were determined using the Euler equations without transversal or friction terms. 
Jung et al.~\cite{Jung2015} developed the approach further, including the relaxation of the transverse terms in the Navier-Stokes equations to be implemented in the isothermal \ac{MRT} \ac{LBM}. Wissocq et al.~\cite{Wiscoq2017} evaluated the non-reflecting boundary condition formulations within the \ac{LBM} for high Reynolds number applications, testing several implementations including the local streamline approach of Albin et al.~\cite{Albin2011} which makes favorable use of the transversal terms. Recently, characteristic boundary conditions were applied within the framework of the fully compressible \ac{LBM} equations~\citep{Gianoli:2023}. Further work by Feng et al.~\citep{feng2019} also explored and evaluated various open boundary treatments, including characteristic approaches, for compressible \ac{LBM} in aerodynamic applications. Compressible \ac{LBM} equations usually involve large velocity stencils, rendering computations expensive in three dimensions~\citep{Wilde2021,Wilde2021a}. 

The \acp{CBC} present a potential building block for \ac{LBM} simulations, extending their use to accurately predict acoustic fields without excessive reflections and improving their ability to resolve turbulent wakes that may interact with outflow boundaries. Since the \ac{LBM} equations are usually formulated for low Mach number flows, the wave amplitudes occurring in the characteristic formulation are different (e.g. zero entropy wave amplitude). In the literature, the incoming wave amplitudes are not directly set to zero but are formulated similarly to a relaxation term to prescribe parameters such as the far-field pressure indirectly, thus preventing it from drifting. The coefficients of the relaxation terms involved are usually set to a constant value on a specific boundary. Additionally, the transversal terms can be adjusted to improve performance for specific flow cases~\citep{Wiscoq2017}. Building on the potential for advanced parameterization and method enhancement, recent research has begun to explore the broader integration of \acp{NN} within \ac{LBM} \citep{bedrunka2021,corbetta2023, horstmann2024}. These neural networks can augment flow simulations in various ways, such as by adding correction values to the numerical method to improve accuracy in simulations of under-resolved flow (e.g., cylinder flow \citep{ataei2024}), or by predicting optimal relaxation parameters to achieve specific, predetermined simulation behaviors \citep{bedrunka2021}.

Inspired by these advancements and the potential for dynamic adaptation, this paper not only focuses on directly reconstructing unknown distribution functions at outflow boundaries but also on determining the model coefficients for the \acp{CBC} using an \ac{NN}. This enables the coefficients to vary \emph{locally} and \emph{temporally}, thus optimizing the effectiveness of the boundary conditions. With this in mind, the primary research question that this paper aims to address is: How can neural networks be leveraged to minimize non-physical reflections from outflow boundaries and preserve aerodynamic accuracy whilst reducing domain length requirements? This is first investigated by directly reconstructing the unknown particle distribution functions at the outflow boundary from the known distributions adjacent to the boundary. Following this, a neural network is used to augment the \acp{CBC} from conventional \ac{CFD} approaches to dynamically tune model coefficients for optimum treatment of outgoing quantities.

This paper is structured as follows. Section~\ref{sec:method} provides an overview of the \ac{LBM}, including an established formulation of standard outflow boundary conditions commonly referred to as the \textit{Zou \& He} model. This section also introduces the details for the simulation design of the benchmark case, flow over a 2D cylinder. Section~\ref{sec:nn_reconstruction} introduces the proposed neural boundary model and details the model training protocols applied on the benchmark case results. In addition, this section presents the direct reconstruction approach and evaluates its performance on a NACA profile. In Section~\ref{sec:nn_cbc}, the neural boundary model is extended to characteristic boundary conditions to improve their stability and physical consistency in open-domain simulations. Finally, Section~\ref{sec:conclusion} summarizes the findings and suggests potential directions for future research.

\section {Methodology} \label{sec:method}

\subsection {The Lattice Boltzmann Method} \label{sec:lbm}
The \ac{LBM} offers a mesoscopic framework for simulating fluid flows, bridging the gap between microscopic particle behavior and macroscopic flow phenomena. This numerical approach models the fluid as an ensemble of particle distribution functions $f_i(\mathbf{x},t)$ that evolve over a discrete lattice. Each distribution function corresponds to a specific discrete velocity direction $\mathbf{c}_i$, and its evolution captures both the streaming (advection) and collision (interaction) phenomena intrinsic to fluid flow. The dynamics of \ac{LBM} are governed by the discrete Boltzmann equation, typically under the single-relaxation-time \ac{BGK} approximation:
\begin{equation}
f_i(\mathbf{x} + \mathbf{c}_i \delta_t, t + \delta_t) - f_i(\mathbf{x}, t) = \Omega_i(\mathbf{f}(\mathbf{x}, t)),
\end{equation}
where $\delta_t$ denotes the time increment and $\mathbf{x}$ the position. The left-hand side represents the streaming step, shifting each distribution to a neighboring node in the direction of $\mathbf{c}_i$, while the right-hand side models local collisions via the operator $\Omega_i$, which assumes a linear relaxation toward a local equilibrium within the \ac{BGK} as 
\begin{equation}
    \Omega_i(f_i) = -\dfrac{1}{\tau}(f_i-f_i^{eq}),
\end{equation}
where $\tau$ is the relaxation time, directly linked to the fluid kinematic viscosity. The equilibrium function $f_i^{eq}$ is described as 
\begin{equation} \label{eq:equilibrium}
    f_{i}^{\text{eq}}\left(\mathbf{x},t\right) = w_{i}\rho\left( 1 + \frac{\mathbf{u \cdot{} c}_{i}}{c_{s}^{2}} + \frac{\left(\mathbf{u \cdot{} c}_{i}\right)^{2}}{2c_{s}^{4}} - \frac{\mathbf{u}\cdot\mathbf{u}}{2c_{s}^{2}} \right),
\end{equation}
with $w_i$ and $c_s$ representing the lattice weights and the speed of sound, respectively. The macroscopic quantities $\rho$ and $\mathbf{u}$ denote the local fluid density and velocity, recovered from the moments of the distribution functions:
\begin{equation} \label{eq:macroscopic_quantities}
    \rho{} \left( \mathbf{x}, t\right) = \sum_{i} f_{i} \left(\mathbf{x},t \right)
    \qquad{} \mathrm{and} \qquad{}
    \rho{} \mathbf{u}\left( \mathbf{x}, t\right) = \sum_{i} \mathbf{c}_{i}f_{i} \left(\mathbf{x},t \right) .
    \vspace{-.3cm}
\end{equation}

\begin{figure}
    \centering
    \includegraphics[width=0.75\linewidth]{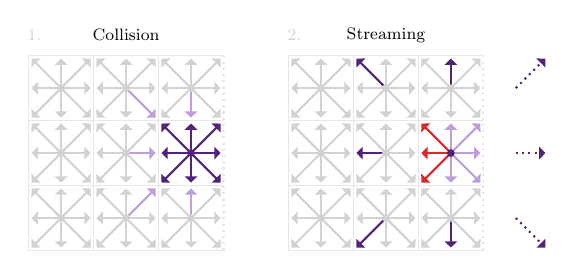}
    \caption{The \ac{LBM} D2Q9 lattice cycle, illustrating (1) collision and subsequent (2) streaming of particle distribution functions. The right \textit{dotted} boundary represents an outflow condition. 
    \textit{Purple} arrows depict the streaming step, while \textit{red} arrows highlight distributions needing reconstruction at this boundary, and \textit{dotted} show distributions that are discarded.}
    \label{fig:lbm_d2q9_cycle}
\end{figure}

Figure~\ref{fig:lbm_d2q9_cycle} illustrates the two essential operations of the \ac{LBM} using a D2Q9 stencil. The figure visually distinguishes the collision step, which acts locally at each lattice node, from the streaming step, where distribution functions propagate along discrete directions (indicated by purple arrows). The right boundary of the domain exemplifies an outflow condition. Here, certain distributions (highlighted in red) must be reconstructed post-streaming to maintain a physically accurate flow field. Distributions that exit the domain and no longer contribute to the solution are shown as dotted arrows.

\subsection {Zou and He Boundary Condition} \label{sec:zou_and_he}
To reconstruct the outward moving distribution functions and recover the desired macroscopic properties on a boundary, the \textit{Zou \& He} boundary condition is a widely adopted approach within the \ac{LBM} \citep{zouhe1997}. 
It enables the enforcement of either the velocity $\mathbf{u} = (u,v) $ or the density $\rho$ and thus the consistent reconstruction of unknown distribution functions (e.g., $f_3, f_6, f_7$ at a right boundary) at the boundary nodes. The central idea is to compute the equilibrium distribution functions (see Eq.~\ref{eq:equilibrium}) using the macroscopic quantities and assuming symmetry of the non-equilibrium components ($f_i^{neq} = f_i-f_i^{eq}$) in opposite directions. Based on this assumption, the unknown distribution functions are reconstructed by combining the computed equilibrium with the non-equilibrium component of their opposite-direction counterparts. Following Eq.~\ref{eq:macroscopic_quantities} the macroscopic quantities are given by:
\begin{equation}
    \rho = f_3 + f_6 + f_7 + \rho_0 + \rho^{+}
\end{equation}
and \vspace{-.3cm}
\begin{equation}
    \rho u = \rho^{+} - (f_1 + f_2 + f_3)
    \vspace{.1cm}
\end{equation}
where $\rho^{+}=f_1+f_5+f_8$  and $\rho_{0}=f_0+f_2+f_4$. This leads to the expression
\begin{equation} \label{eq:rho_u_relation}
    \rho = \dfrac{1}{1+u}(\rho_0 + 2 \rho^+)
\end{equation}
which establishes a direct relationship between the velocity and density at the boundary. This formulation corresponds to a Dirichlet boundary condition, where one macroscopic quantity (either $\rho$ or $u$) is prescribed, and the other is subsequently inferred from the local known distribution functions. This leads to the following expressions for the reconstructed distributions at a right boundary:\vspace{-.1cm}
\begin{align}
    g_6 &= f_6^{eq} + f_8^{neq} + \dfrac{1}{2}\left( f_4^{neq}-f_2^{neq}\right)\\
    g_3 &= f_3^{eq} + f_1^{neq}\\
    g_7 &= f_7^{eq} + f_5^{neq} - \dfrac{1}{2}\left( f_4^{neq}-f_2^{neq}\right)
\end{align}

The other distributions are kept unchanged so that
\begin{equation}
g_i = f_i \quad \text{for}~~ i=0,1,2,4,5,8
\end{equation}

This approach can be extended to the top, bottom, and left boundaries by symmetry, modifying the directional indices accordingly.

\subsection {Preliminary Benchmark for Flow Over a 2D Cylinder}
\begin{figure}
    \centering
   \includegraphics[height=0.3\linewidth]{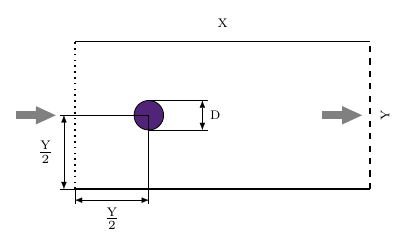}
   \vspace{-.4cm}
    \caption{Domain geometry: The configuration of the cylinder application is defined by the number of grid points across the diameter \textsc{d}. The cylinder diameter \textsc{d} is placed at $\frac{\textsc{y}}{2}$ from the inlet. The stream-wise domain length \textsc{x} and the domain height \textsc{y} are multiple of the cylinder diameter \textsc{d}, respectively. Boundary conditions: Inlet on the left (dotted line), outlet on the right (dashed line), periodic boundaries for all lateral boundaries.}
    \label{fig:cylinder2d}
\end{figure}

The flow around a cylinder represents a classical test case for fluid-structure interactions. This scenario has been the subject of numerous experimental and numerical studies over decades and remains an established benchmark due to its simplicity, relevance, and applicability to more complex scenarios. The downstream flow is characterized by a distinctive Kármán vortex street and the surrounding laminar flow. Due to the variety of flow phenomena within the domain, the two-dimensional cylinder flow serves as a training set to train the neural boundary model. The geometry of the computational domain is shown in Figure~\ref{fig:cylinder2d}. The streamwise length \textsc{x} of the domain is determined by scaling the cylinder's diameter \textsc{d} with a specific factor in the stream-wise direction, while the domain height \textsc{y} is determined by scaling the diameter \textsc{d} with a corresponding factor in the transverse direction. The center of the cylinder is located at $\mathbf{x}=(\textsc{y}/2,\textsc{y}/2)$. The boundary of the solid is modeled through the full-way bounce-back method, which enforces the no-slip condition at the fluid-solid interface. In this approach distributions impinging on the solid surface are reflected back along their incoming directions as
\begin{equation}
f_{\bar{i}}(\boldsymbol{x}, t + \Delta t) = f_i^{\star}(\boldsymbol{x}, t - \Delta t),
\end{equation}
effectively reversing their momentum. $f_i^{\star}$ denotes the post-collision distribution function and $f_{\bar{i}}$ the distribution function of their counterparts with respect to $\boldsymbol{c}_i$. 
The benchmark evaluates aerodynamic performance by comparing force coefficients (i.e., drag $C_D$ and lift $C_L$) as well as the Strouhal number against reference data from literature. These quantities are calculated by
\begin{equation}\label{eq:drag_lift}
    C_D=\frac{2F_x}{\rho U^2A} \quad \text{and} \quad C_L=\frac{2F_y}{\rho U^2A},
\end{equation}

\begin{figure}[!b]
    \centering
    \includegraphics[width=0.32\linewidth]{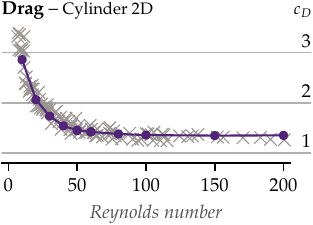}
    \hfill
    \includegraphics[width=0.32\linewidth]{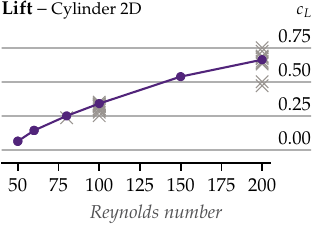}
    \hfill
    \includegraphics[width=0.32\linewidth]{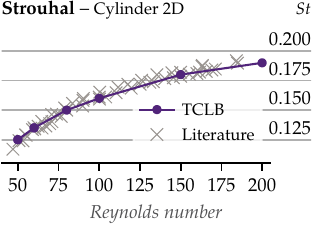}
    \caption{Benchmark evaluation of the training set configuration, comparing Drag, Lift, and Strouhal numbers with values from the literature. The configuration parameters are set to $\textsc{D}=50$, $X=100\times\textsc{D}$, $Y=50\times\textsc{D}$, and $\textit{Mach}=0.1$. Literature is taken from \cite{tritton1959, weiselsberger1922, hoerner1992, roshko1953, norberg2003, williamson1989, shu2007}}
    \label{fig:cylinder_benchmark}
\end{figure}

where $F_x$ and $F_y$ denote the fluid-induced forces acting on the cylinder in streamwise and transverse directions, respectively. Here, $U$ describes the characteristic velocity and $A$ the projected area of the body perpendicular to the corresponding force direction. In \ac{LBM}, the momentum exchange method is a widely used and efficient approach for calculating the hydrodynamic force. In the context of bounce-back boundary conditions for stationary geometries, the total momentum exchanged at the fluid-solid interface is computed by summing the contributions from the incoming and reflected populations at the boundary as
\begin{equation}\label{eq:MEM_basic}
    \Delta P= \Delta x^3\sum_{x_i^b}(f_i^{in}\boldsymbol{c}_i-f_{\Bar{i}}^{out}\boldsymbol{c}_{\Bar{i}}).
\end{equation}
where $f$ denotes the incoming distributions and $f$ the reflected distributions. Finally, the momentum exchange $\Delta P$ directly yields the force $\mathbf{F}$ acting on the object by dividing by the time step:\vspace{-.25cm}
\begin{equation}
    \boldsymbol{F}={\Delta P}/{\Delta t}.
    \vspace{.15cm}
\end{equation}
The Strouhal number is then captured by a fast Fourier transformation of the lift coefficient. Figure~\ref{fig:cylinder_benchmark} shows the force coefficients for Reynolds numbers in a range between 10 and 200 for a domain configuration of $\textsc{D}=50$, $X=100\times\textsc{D}$, $Y=50\times\textsc{D}$. The simulations to generate the results and later the training datasets for the direct distribution reconstruction approach were performed using the in-house \textit{\textsc{TCBL}} software package \citep{tclb}. As shown in the figure, the results are in good agreement with the literature that can be used for the training procedure and evaluation. 

\section {Direct Distribution Reconstruction} \label{sec:nn_reconstruction}
\subsection {Neural Network Boundary Condition for Flow Over a 2D Cylinder}

\begin{figure}[!b]
\textit{\scriptsize{\hspace{1.0cm}A. \hspace{4.8cm} B.}}\par\medskip\vspace{-.5cm}
    \centering
    \includegraphics[height=0.3\linewidth, angle=90]{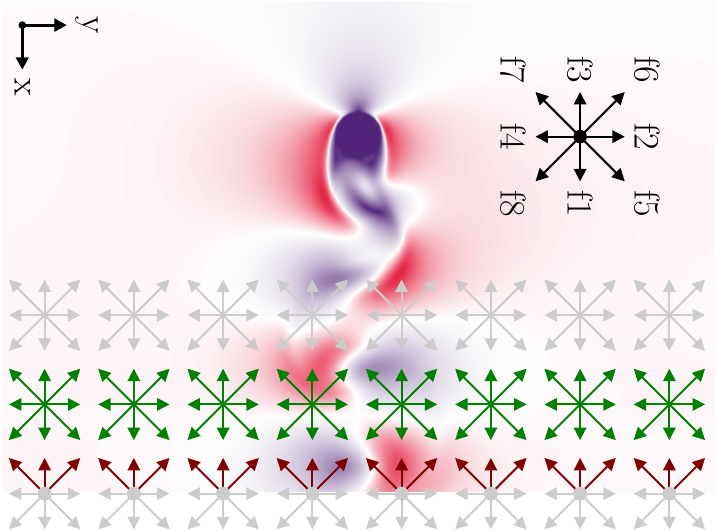}
    \hspace{1cm}
    \includegraphics[height=0.4\linewidth]{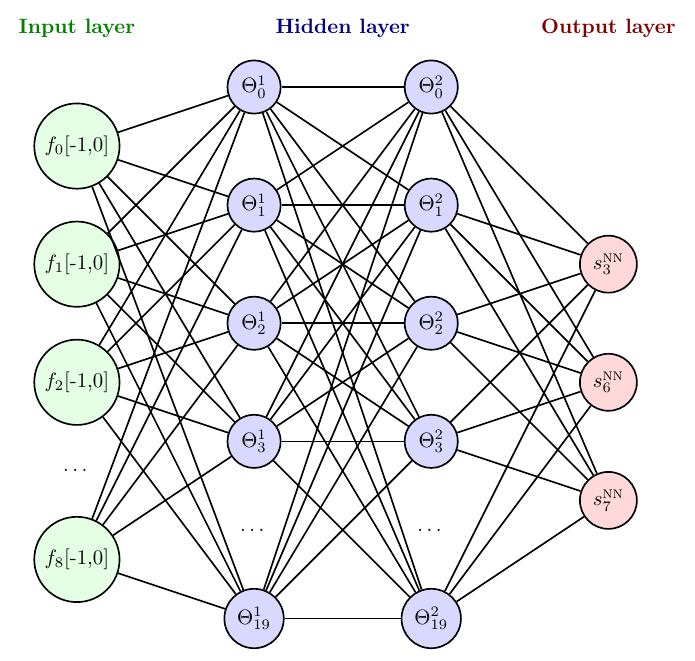}
    \caption{A. Procedure; B. Illustration of the neural network topology, including the standard network with two hidden layer.}
    \label{fig:neural-network_procedure}
\end{figure}

\begin{figure}
    \textit{\scriptsize{\hspace{-.2cm}A. \hspace{2.6cm} B.\hspace{2.6cm} C.}}\par\medskip
    \centering
    \includegraphics[height=0.32\linewidth]{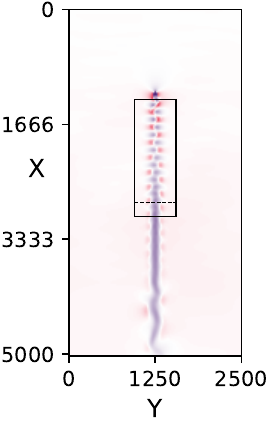}
    \includegraphics[height=0.32\linewidth]{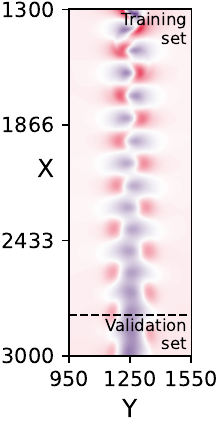}
    \includegraphics[height=0.32\linewidth]{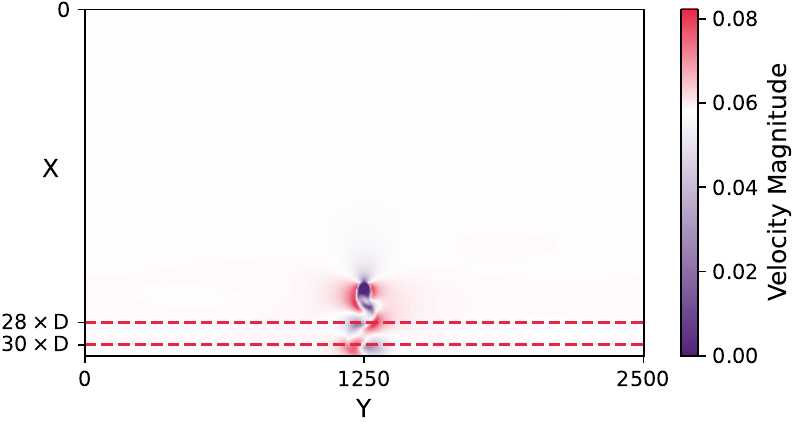}
    \caption{\textit{A.} Overview of the large benchmark domain with $\textit{D}=50$, $X=100\times\textsc{D}$, $Y=50\times\textsc{D}$, configured to minimize the influence of the outflow boundary condition on the obstacle. The configuration shows the $\textit{Re}=200, \textit{Mach}=0.1$ case. The center of the cylinder is located at $25 \times\textsc{D}$. The rectangular region highlights the area used for training and validation. \textit{B.} Visualization of the training and validation datasets employed to train the neural boundary condition. \textit{C.} Reduced evaluation domain, cut at the dashed lines, used to assess the performance of different outflow boundary conditions in the wake region behind the obstacle.}
    \label{fig:training}
\end{figure}

The first approach proposed in this work uses a neural network to reconstruct the unknown distribution functions at the outflow boundary. This section investigates the capability of this method to recover local flow features and quantifies the benefits this can bring in significantly reducing the influence of the boundary on internal aerodynamic metrics (i.e. lift, drag and Strouhal number). 
The aim is to infer appropriate corrections that mitigate boundary-induced artifacts without requiring information beyond the interior of the domain. 
As illustrated in Figure~\ref{fig:neural-network_procedure}, the reconstruction process is based on a shallow neural network architecture. The input consists of the nine local distribution functions at the penultimate lattice layer $x=\textsc{L}-1$. The network architecture comprises two hidden layers with 20 neurons each and outputs a correction term $s_i^{\textsc{nn}}$, which is added to the corresponding distributions to yield the modified values:\vspace{-.2cm}
\begin{equation}
    f_i^{\textsc{b}}(x=\textsc{L}) = f_i(x=\textsc{L}-1)+s_i^{\textsc{nn}}  \quad \text{for}~~ i=3,6,7.
    \vspace{.2cm}
\end{equation}
The idea is that this approach allows the model to learn physically consistent boundary behavior from interior flow characteristics, without explicitly encoding boundary physics. 
For this purpose, the neural network is trained on a snapshot of a converged cylinder flow at Reynolds number 200 and Mach number 0.1, shown in Figure~\ref{fig:training}. The snapshot includes both the characteristic Kármán vortex street in the wake region and laminar flow zones near the domain edges. From this dataset, separate training and validation sets are constructed to ensure generalization capability. To train the neural boundary model, a supervised learning strategy is employed, in which the network learns to predict the distribution functions in one layer downstream based on those of the preceding layer. Specifically, distribution data from position $x-1$ are provided as input to the network, while the corresponding values at $x$ serve as the ground truth. Each layer of the selected training domain is treated as a single batch, resulting in a total of 1500 batches. During training, batches are randomly shuffled to promote generalization and prevent overfitting. The training objective is defined through the mean squared error (MSE) between the predicted distribution functions $f_i^{\textsc{b}}$ and the reference distribution functions $f_i^{\textsc{ref}}$:
\begin{equation}
\vspace{-.2cm}
    \mathcal{L} = \text{MSE}\left(f_i^{\textsc{b}}(x), f_i^{\text{ref}}(x)\right), \quad \text{for} \quad i = 3, 6, 7.
\end{equation}

The model is trained over 200 epochs using a learning rate of $1\times10^{-5}$. Convergence is observed in both the training and validation sets, with no indication of overfitting, confirming that the network generalizes well within the selected snapshot. To assess the effectiveness of the trained boundary model, the computational domain is truncated at various downstream locations behind the cylinder (illustrated as red dotted lines in Figure~\ref{fig:training}.C. This evaluation strategy aims to determine whether the neural network–based outflow boundary can preserve the dynamics of the flow field more accurately than traditional approaches, such as the \textit{Zou \& He} boundary condition. The evaluation of the proposed boundary condition is conducted over a range of Reynolds numbers from 10 to 200. To assess its sensitivity to numerical parameters, simulations are performed under both diffusive and acoustic scaling. This allows investigation of the effects of varying the relaxation times (acoustic scaling) and Mach numbers (diffusive scaling) relative to the conditions used during training.
\begin{figure}
    \centering
    \vspace{0.5cm}
    \includegraphics[width=0.49\linewidth]{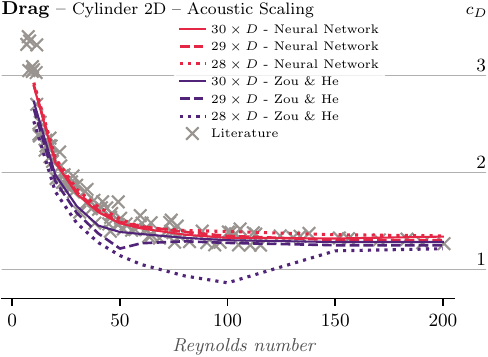}
    \includegraphics[width=0.49\linewidth]{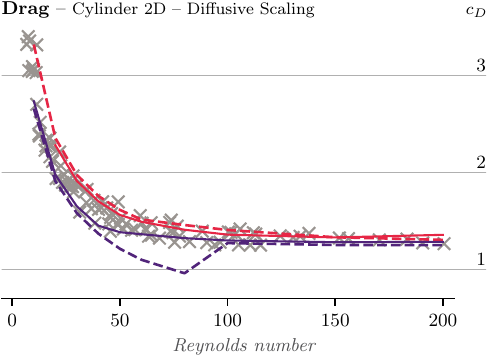}
    \\
    \vspace{0.5cm}
    \includegraphics[width=0.49\linewidth]{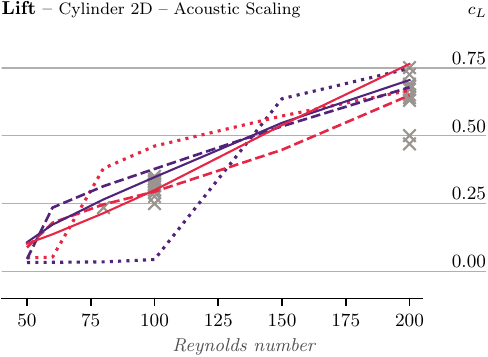}
    \includegraphics[width=0.49\linewidth]{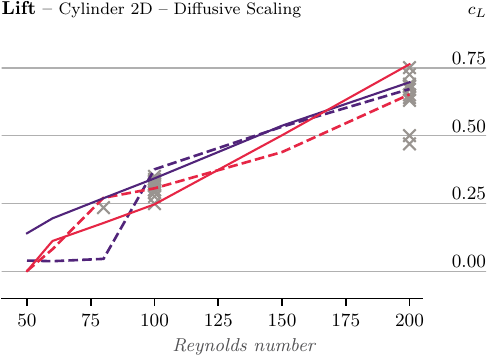}
    \\
    \vspace{0.5cm}
    \includegraphics[width=0.49\linewidth]{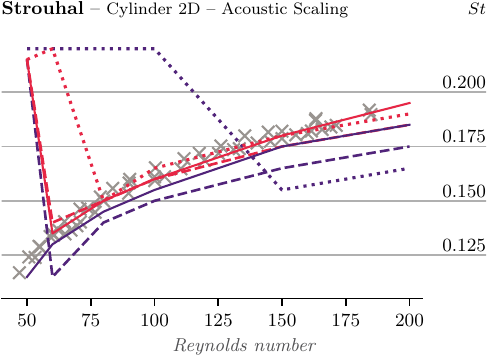}
    \includegraphics[width=0.49\linewidth]{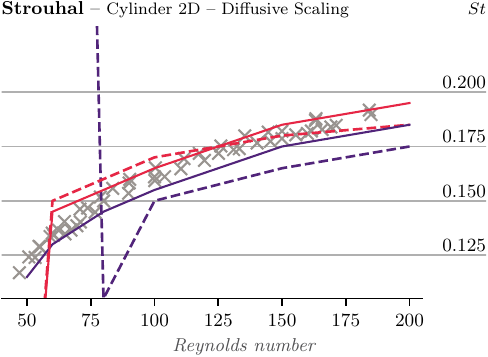}
    \caption{Evaluation of the neural boundary condition (red lines) and the \textit{Zou \& He} (purple lines) for the two-dimensional cylinder application, comparing Drag ($C_D$), Lift ($C_L$), and Strouhal ($St$) numbers with values from the literature (e.g., grey crosses). The configuration parameters are set to $\textit{D}=50$, $Y=50\times\textsc{D}$, and $\textit{Mach}=0.1$ using \ac{BGK}. Results are presented for different outflow boundary locations, effectively varying the simulated wake region length: $30\times D$ (line), $29\times D$ (broken line), and $28\times D$ (dotted line), where $D$ is the cylinder diameter. Literature is taken from \cite{tritton1959, weiselsberger1922, hoerner1992, roshko1953, norberg2003, williamson1989, shu2007}}
    \label{fig:cylinder2d_results}
\end{figure}
Figure~\ref{fig:cylinder2d_results} presents the results for the aerodynamic force coefficients (drag $C_D$, lift $C_L$) and the Strouhal number $St$. It is evident that the reconstructed distribution functions provided by the neural network yield significantly better agreement with the reference data than the classical \textit{Zou \& He} boundary condition. This advantage is observed in both acoustic and diffusive scaling. Even with a substantial reduction in domain length, the learned boundary model is able to reliably reproduce the characteristic flow structures and key physical quantities. A comparison of the two scaling approaches further reveals that the neural model remains robust when applied to configurations with varying viscosity at a constant Mach number. However, under diffusive scaling, increasing the deviation from the training Reynolds number leads to a corresponding rise in Mach number. In these cases, noticeable deviations from the literature values begin to emerge. However, the predictive accuracy of the neural boundary model continues to outperform that of the \textit{Zou \& He} condition. In contrast, the \textit{Zou \& He} boundary condition exhibits noticeable deviations from the literature values. A key observation is that at lower Reynolds numbers, the \textit{Zou \& He} method is often incapable of accurately representing the expected values for drag and lift. For instance, in Figure~\ref{fig:cylinder2d_results}, the \textit{Zou \& He} results (purple lines) consistently underestimate the drag coefficient and Strouhal number compared to both the literature and the \ac{NN} model across various Reynolds numbers and truncations. While there's an observable trend that the \textit{Zou \& He} results tend to improve slightly as the domain length increases (i.e., with a longer wake region, comparing results for $30\times D$ versus $28\times D$ in acoustic scaling), the numerical influence of this boundary condition remains significant.

\subsection {Application of Neural Network Boundary Condition to NACA0012 airfoil }

\begin{figure}
    \centering
    \vspace{0.5cm}
    \includegraphics[width=0.75\linewidth]{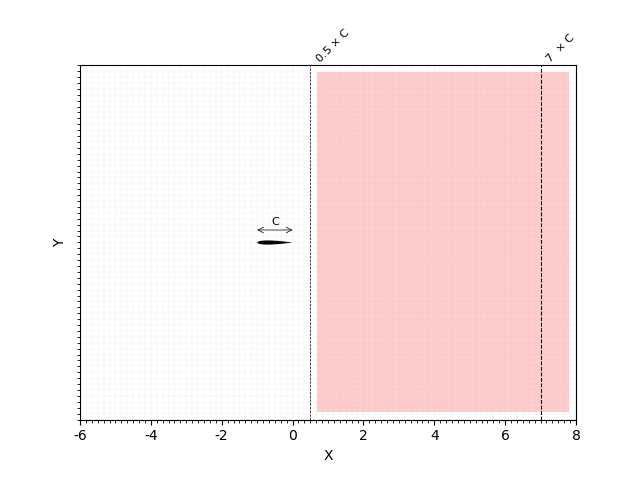}\\
    \vspace{-.3cm}
    \caption{Truncated computational domain used for the \textit{NACA0012} airfoil simulations to assess outflow boundary condition performance. The shaded area represents the truncated wake region, whose dimensions are proportional to the airfoil chord length C.} 
    \label{fig:naca2d_configuration}
\end{figure}

To assess the generalizability of the trained neural network boundary condition, the previously trained model is directly applied to study flow over a \textit{NACA0012} airfoil. See Appendix~\ref{Appendix:Naca0012} for details on the airfoil geometry generation. This presents an alternative geometric configuration to highlight the robustness of the boundary condition. The simulation conditions are largely comparable to those used for the cylinder case; however, to improve numerical stability, the \ac{MRT} collision operator is employed instead of the \ac{BGK} model. The simulations are carried out with a Reynolds number of $Re=1000$. The evaluation begins with a reference simulation featuring an extended wake region, which yields results that are in good agreement with the data from the literature. Subsequently, the domain is significantly truncated, positioning the outflow boundary only 0.5 chord lengths $C$ downstream of the airfoil trailing edge as illustrated in Figure~\ref{fig:naca2d_configuration}.

\begin{figure}
    \centering
    \vspace{0.5cm}
    \includegraphics[width=0.49\linewidth]{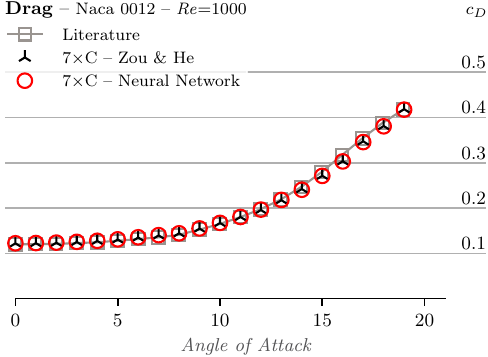}
    \includegraphics[width=0.49\linewidth]{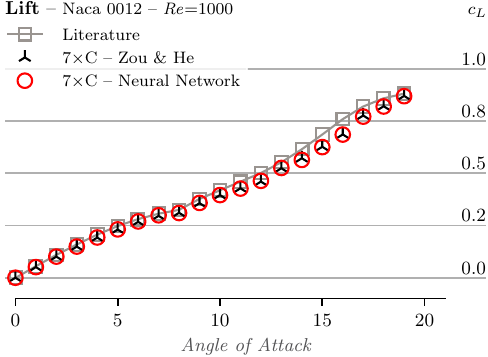}\\[.4cm]
    \includegraphics[width=0.49\linewidth]{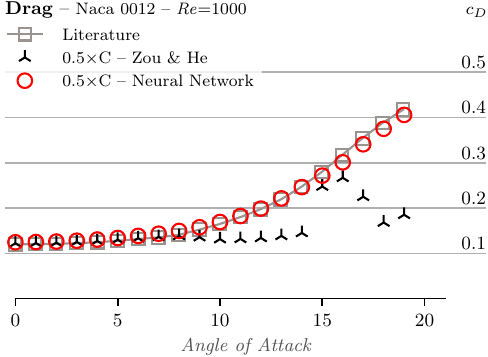}
    \includegraphics[width=0.49\linewidth]{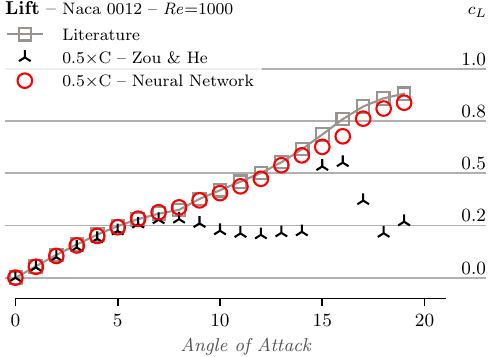}
    \caption{Evaluation of the neural boundary condition, comparing Drag, and Lift numbers with values from the literature and the \textit{Zou \& He} model for the two dimensional \textit{NACA0012} application. Two cases are shown, representing domain lengths ending $7C$ (top row) and $0.5C$ (bottom row) after the trailing edge, respectively, with chord length $C$. Reference is taken from~\cite{kurtulus2015}.} 
    \label{fig:naca2d}
\end{figure}

In this configuration, the drag and lift coefficients are analyzed across a range of angles of attack, see Figure \ref{fig:naca2d} for results. As the angle of attack increases, so does the flow complexity as a result of the emergence of stronger coherent vortex structures, placing higher demands on the boundary condition. The results demonstrate that the neural boundary model is substantially more robust in handling outgoing vortices through improved reconstruction of missing distribution functions compared to the classical \textit{Zou \& He} condition and data from the literature \citep{kurtulus2015}. The results of applying the \textit{Zou \& He} condition on the truncated domain start to deviate from the literature above an angle of attack of $7^\circ$. At these higher angles of attack, the wake behavior strongly influences the aerodynamic force coefficients but is not resolved by the \textit{Zou \& He} condition in the truncated domain. The ability to accurately predict flow characteristics and time-averaged aerodynamic force coefficients despite a shortened domain highlights the potential of the neural distribution reconstruction method to reduce computational cost without affecting the physical fidelity of the simulation.

\section{Neural Extended Characteristic Boundary Conditions} \label{sec:nn_cbc}
This study also explores the integration of neural networks into established outflow boundary models to address the issue of pressure wave reflections. The goal is to improve the stability and physical consistency of simulations, particularly those focused on capturing acoustic phenomena. For this purpose, an adaptively designed neural network is developed, capable of dynamically adjusting tuning parameters in both space and time based on local flow data, to achieve an optimized boundary treatment.

\acp{CBC} provide boundary conditions based on the characteristic form of the Euler or Navier-Stokes  equations. Unlike traditional Dirichlet or Neumann conditions, \acp{CBC} are particularly effective in minimizing artificial reflections at domain boundaries. 
This is achieved by diagonalizing the flux terms for directions normal to the respective boundary plane individually \citep{Poinsot:1992,Lodato2008,LBMCBC:2008,Wiscoq2017}. The eigenvalues characterize the propagation directions of the characteristic wave components, thus indicating incoming or outgoing information. They are part of the amplitude time variations of the characteristic waves, $\mathcal{L}$, introduced during the derivation process \citep{Lodato2008}. Those amplitudes can be used to rewrite the Navier-Stokes equations in a characteristic-type form, suitable to prescribe boundary conditions to treat incoming and outgoing wave components consistently according to their physical propagation direction.
Details for deriving the characteristic boundary conditions can be found in the literature cited above. The governing equations are summarized in the following for a simple two-dimensional problem for a propagation direction in the positive x-direction:
\begin{equation} \label{eq:rho_dt}
    \dfrac{\partial \rho}{\partial t} + \dfrac{1}{2 c_s^2} \left( \mathcal{L}_5 + \mathcal{L}_1\right) = \dfrac{1}{2 c_s^2} \left( \mathcal{T}_5 + \mathcal{T}_1\right)
\end{equation}
\begin{equation}\label{eq:u_dt}
    \dfrac{\partial u}{\partial t} + \dfrac{1}{2 \rho c_s} \left( \mathcal{L}_5 - \mathcal{L}_1\right) = \dfrac{1}{2 \rho c_s} \left( \mathcal{T}_5 - \mathcal{T}_1\right)
\end{equation}
\begin{equation}\label{eq:v_dt}
    \dfrac{\partial v}{\partial t} + \mathcal{L}_3 = \mathcal{T}_3.
\end{equation}
Here, the wave amplitude terms are usually defined as
\begin{align}
    \mathcal{L}_1 &= (u-c) \left( \dfrac{\partial \rho}{\partial x} - \rho c_s \dfrac{\partial u}{\partial x} \right)\\
    \mathcal{L}_3 &= u \dfrac{\partial v}{\partial x}\\
    \mathcal{L}_5 &= (u+c) \left( \dfrac{\partial \rho}{\partial x} + \rho c_s \dfrac{\partial u}{\partial x} \right),
\end{align}
with the entropy wave amplitude $ 
    \mathcal{L}_2 = u \left( c_s^2\dfrac{\partial \rho}{\partial x} - \dfrac{\partial p}{\partial x} \right)$ being zero \citep{Wiscoq2017}.
The transversal components $\mathcal{T}$ are defined as
\begin{align}
    \mathcal{T}_1 &= - \left[ v \dfrac{\partial p}{\partial y} + p \dfrac{\partial v}{\partial y} - \rho c_s v \dfrac{\partial v}{\partial y}\right]\\
    \mathcal{T}_3 &= - \left[ v \dfrac{\partial v}{\partial y} + \dfrac{1}{\rho} \dfrac{\partial p}{\partial y} \right]\\
    \mathcal{T}_5 &= - \left[ v \dfrac{\partial p}{\partial y} + p \dfrac{\partial v}{\partial y} + \rho c_s v \dfrac{\partial v}{\partial y}\right].
\end{align}
The terms $\mathcal{L}_4, \mathcal{T}_2$  and $\mathcal{T}_4$ are not introduced, here, as they vanish in an isothermal and two-dimensional domain, respectively. 
To avoid reflections at the outflow boundary, the incoming wave amplitude $\mathcal{L}_1$ is adjusted according to Wissocq et al.~\cite{Wiscoq2017} in the following way
\begin{equation} \label{eq:cbc_L1}
    \mathcal{L}_1 = \kappa_{1} (p-p_\infty) - \kappa_2 (\mathcal{T}_1-\mathcal{T}_{1, exact}) + \mathcal{T}_1
\end{equation}
with 
\begin{equation}
    \kappa_{1} = \sigma (1-M^{2}) c_s / L.
\end{equation}
Here, $\sigma$ and $\kappa_{2}$ are free parameters to be discussed below, $M$ the Mach number and $L$ a characteristic length scale.
Once the time derivatives of the macroscopic quantities (Eqs.~\ref{eq:rho_dt}, \ref{eq:u_dt} and \ref{eq:v_dt}) are evaluated, a first-order Euler scheme is employed to prescribe these variables for the next time step.
To map the updated macroscopic variables back into distribution space, the \textit{Zou \& He} reconstruction algorithm is applied (see Section~\ref{sec:zou_and_he}). However, since Eq.~\ref{eq:rho_u_relation} establishes a direct relationship between velocity and density at the boundary, and both are already determined via the known distribution functions, the system becomes over-constrained. In this case, the only remaining degree of freedom lies in adjusting the distribution functions themselves. To resolve this, Wissocq et al.~\cite{Wiscoq2017} propose introducing a correction term to the the zero-velocity distribution function $f_0$ to maintain consistency with the prescribed macroscopic state. The corrected expression reads:
\begin{equation}
    g_0 = f_0 + \rho - \dfrac{1}{1+u}\left( \rho_0 + 2 \rho_{+} \right)
\end{equation}

\subsection{Benchmark Description for a 2D Convected Vortex}

A well-established benchmark for evaluating the performance of non-reflecting boundary conditions is the two-dimensional convected vortex, in which a vortex is advected across the computational domain and exits through the outflow boundary \citep{Poinsot:1992,Jung2015}. Following the setup described by Wissocq et al.~\cite{Wiscoq2017}, we adopt the isothermal initialization of a Lamb–Oseen vortex, 
\begin{align}
u &= u_0 - \beta u_0 \dfrac{(y-y_0)}{R_c}\text{exp}\left(-\dfrac{r^2}{2 R_c}\right), \\
v &= \beta u_0 \dfrac{(x-x_0)}{R_c}\text{exp}\left(-\dfrac{r^2}{2 R_c}\right), \\
\rho &= \left[ 1- \dfrac{(\beta u_0)^2}{2 C_v} \text{exp}\left(-\dfrac{r^2}{2}\right)\right]^{1/(\gamma-1)}~.
    \vspace{0.2cm}
\end{align}

The initial velocity field $\mathbf{u}$ and density $\rho$ are prescribed analytically, with the center of the vortex positioned at $x_0=y_0=150$ on a $200\times200$ lattice domain. The reference velocity is defined as $u_0 = Ma \cdot c_s$, where the Mach number is set to 0.3 and the isothermal speed of sound $c_s$ is consistent with the \ac{LBM} formulation. The parameter $\gamma =2$, core radius $R_c=20$, and scaling factor $\beta = 0.5$ are chosen to control the vortex strength and compactness. To mitigate artificial reflections, the domain is extended in both the streamwise and crosswise directions. This ensures that waves reflected from the inlet and propagating through the periodic lateral boundaries do not interfere with the central region, where the loss function is evaluated. The Reynolds number, based on $u_0$ and the domain length, is set to 750. All simulations use the \ac{BGK} collision model. The simulations for the two-dimensional convected vortex, as well as the development and time-dependent training of the neural network for the extended \ac{CBC}, were carried out using the \ac{LBM} framework \textit{\textsc{lettuce}}. The neural network components were implemented and trained leveraging \textit{\textsc{lettuce}}'s \textsc{PyTorch} backend \citep{bedrunka2021}.

\subsection{Neural Networks for Tuning Parameter}
This study explores a neural network, to dynamically adapt the parameters of the \ac{CBC} and minimize spurious reflections. This trained neural network processes nine input features for each node along the outflow boundary. These features comprise local spatial derivatives of the pressure and velocity components in both x and y directions ($p_x$, $u_x$, $v_x$, $p_y$, $u_y$, $_vy$), as well as temporal changes in the pressure (or density) and velocity components ($p_t$,$u_t$,$v_t$) from the preceding time step, which are already calculated for processing the \ac{CBC} model. The network architecture consists of two hidden layers, with 20 neurons each, connected through a LeakyReLU activation function. This results in an output layer that produces two scalar values $s_0$ and $s_1$. These two scalar outputs from the network undergo a post-processing step. Each output is first passed through a Sigmoid activation function. Subsequently, these activated values are scaled and shifted using predetermined coefficients to yield two final tuning parameters, $\sigma$ and $\kappa_2$. These parameters are then used to adaptively adjust the coefficients in Eq.~\ref{eq:cbc_L1} within the expression for the incoming characteristic wave amplitude, $\mathcal{L}_1$, of the \ac{CBC}.

\begin{figure}
    \centering
    \vspace{0.5cm}
    \includegraphics[width=0.99\linewidth]{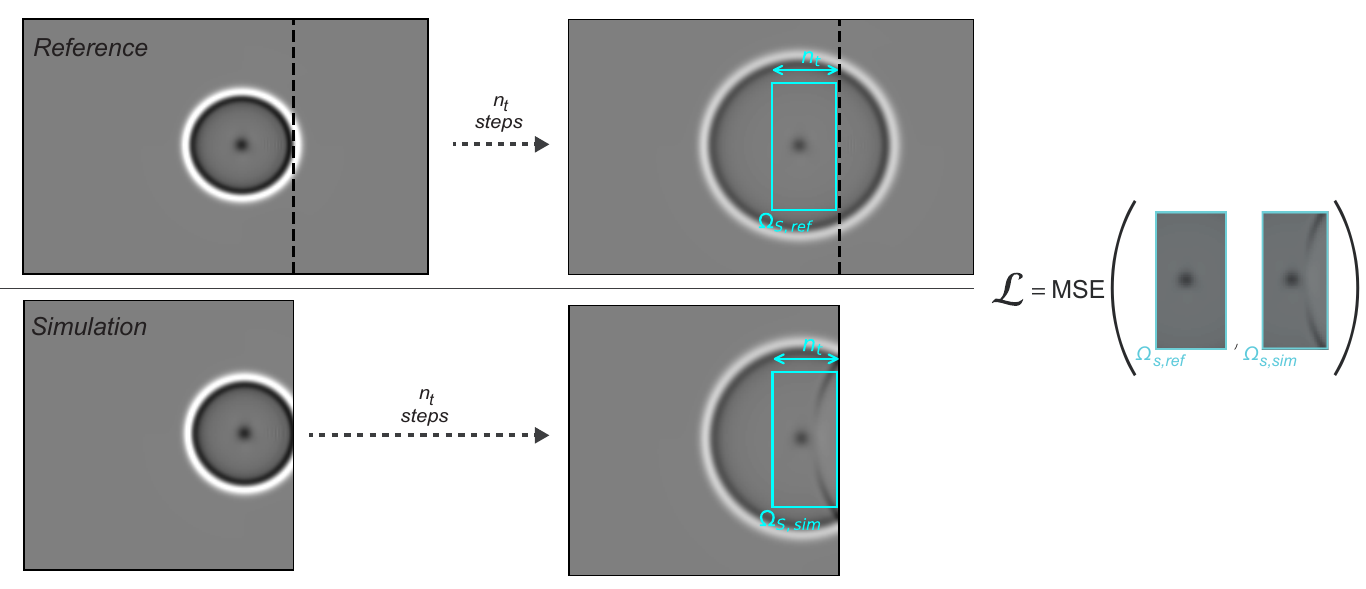}
    \caption{Illustration of the training process for the neural network-tuned Characteristic Boundary Conditions (CBC). A short simulation segment is initialized from a reference state and run for $n_t$ steps using the neural network predicted \ac{CBC} parameters. The resulting fields (here density) within the sub-region $\Omega_{S,sim}$ are then compared against the corresponding data from the reflection-free reference simulation $\Omega_{S,ref}$ to compute the loss $\mathcal{L}$.}
    \label{fig:tuningProcedure} 
\end{figure}

The neural network for tuning the \ac{CBC} parameters ($\sigma$ and $\kappa_2$) is trained by minimizing a loss function based on the Mean Squared Error (MSE). This loss quantifies the discrepancy between the simulated density $\rho_{sim}$ and velocity $\mathbf{U}_{sim}$ fields and the target reference fields ($\rho_{ref}$ and $\mathbf{U}_{ref}$). The reference data were sourced from pre-computed simulations of the 2D convected vortex benchmark on an extended domain to minimize boundary influences. Figure~\ref{fig:tuningProcedure} illustrates the training procedure. For each short simulation segment used during training, the \ac{LBM} simulation is run for $n_t$ time steps using the \ac{CBC} parameters provided by the neural network. At the end of this segment (i.e., at time t=$n_t$), the instantaneous density field $\rho_{sim}(x,y,t=n_t)$ and velocity field $\mathbf{U}_{sim}(x,y,t=n_t)$) are recorded. The loss for a single such segment is then calculated as the sum of the MSE for density and velocity fields. The MSE is calculated over a dynamically defined spatial sub-region $\Omega_s$ within the computational domain, whose extent depends on the segment duration $n_t$. This sub-region covers the area potentially influenced by boundary reflections. Assuming that the outflow boundary is located at $x=L$ (the last x-coordinate of the computational domain), $\Omega_s(n_t)$ includes all grid cells (x,y) such that  $L-n_t < x < L$ and for all y values that span the height of the domain. This ensures that for a segment of $n_t$ steps, the loss is computed over the $n_t$ layers closest to the outflow boundary. The loss $\mathcal{L}$ is thus defined as:
\begin{equation}
    \mathcal{L} = \text{MSE}_{\Omega_s}\left(\rho_{sim}(n_t), \rho_{ref}(n_t)\right) + \text{MSE}_{\Omega_s}\left(\mathbf{U}_{sim}(n_t), \mathbf{U}_{ref}(n_t)\right),
\end{equation}
where $MSE_{\Omega_s}\left(A,B\right) = \frac{1}{\vert \Omega_s \vert} \sum_{(x,y) \in \Omega_s} \left(A(x,y)-B(x,y) \right)^2$ and $\vert \Omega_s \vert$ denotes the number of grid points within the sub-region $\Omega_s$. This approach evaluates the performance of the \ac{NN}-tuned \ac{CBC} parameters at a specific point in the evolution of each short simulation. This allows the network to learn parameters that minimize deviations from the reflection-free reference solution.

From the reflection-free reference simulation, 110 distinct initial flow fields were extracted for the training data. These fields were captured at intervals of five time steps, starting from time step 1 up to time step 550 of the reference simulation. Each of these saved fields served as an initial condition for subsequent short simulation segments used in the training process. To expose the neural network to a variety of scenarios and time horizons, each of these 110 initial fields was used to launch simulation segments of varying durations. Specifically, the \textit{Neural Network} was trained using segments with $n_t\in{5,10,25}$ time steps. Further training configurations of the network utilizing extended time segments such as $n_t\in{5,10,25,50}$ time steps, leading to the same behavior as presented in this study. At the end of each of these respective durations, the loss function was evaluated by comparing the state of the segment.

\begin{figure}
    \centering
    \newlength{\imgwidth}
    \setlength{\imgwidth}{0.1385\linewidth} 

    \includegraphics[width=4\imgwidth]{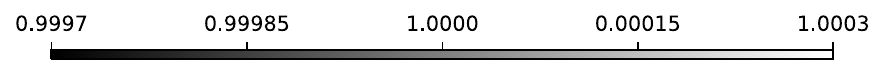}

    \parbox{\linewidth}{\centering\textbf{Reference}}\par\vspace{1pt} 
    \includegraphics[width=\imgwidth]{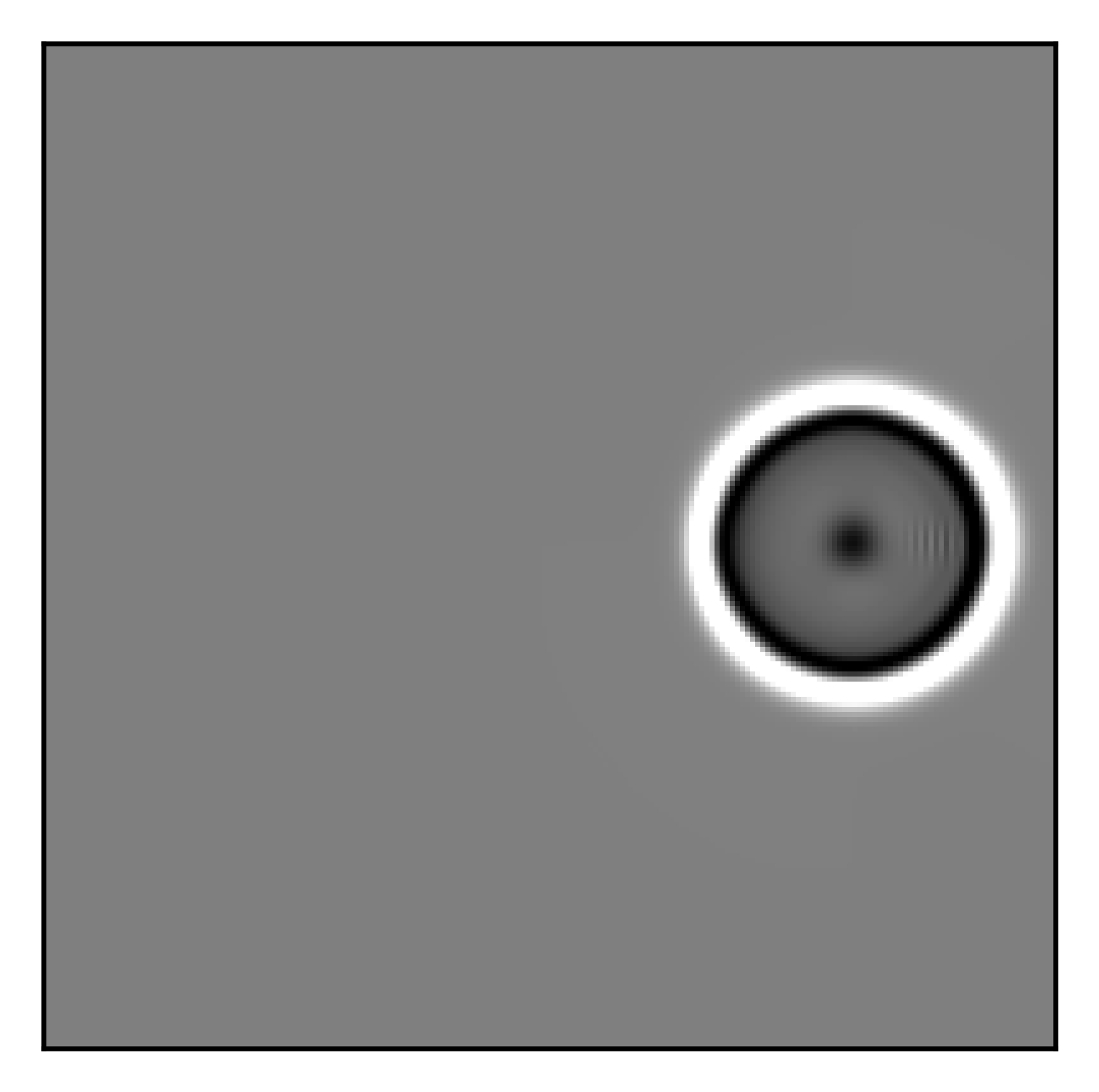}
    \includegraphics[width=\imgwidth]{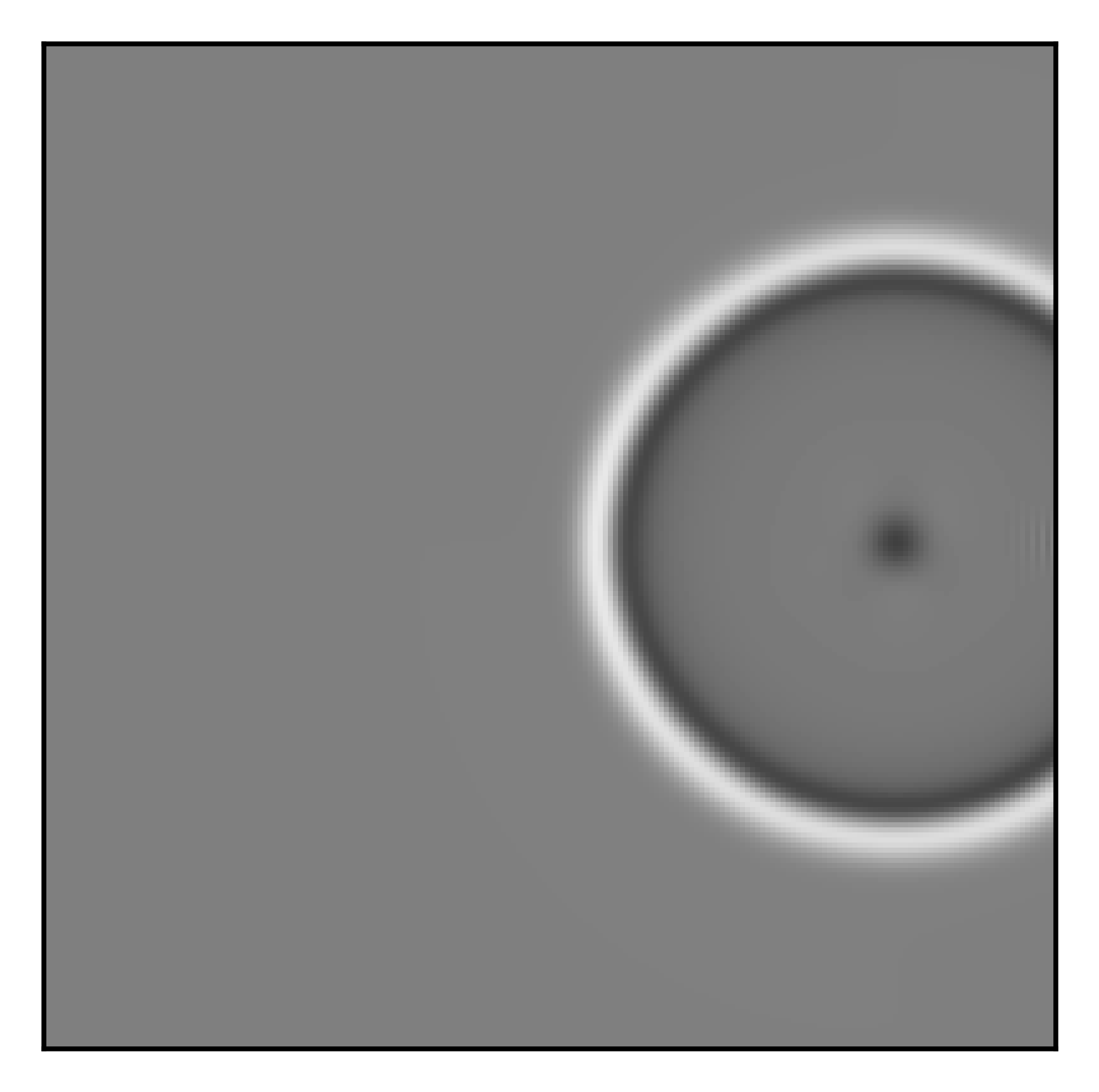}
    \includegraphics[width=\imgwidth]{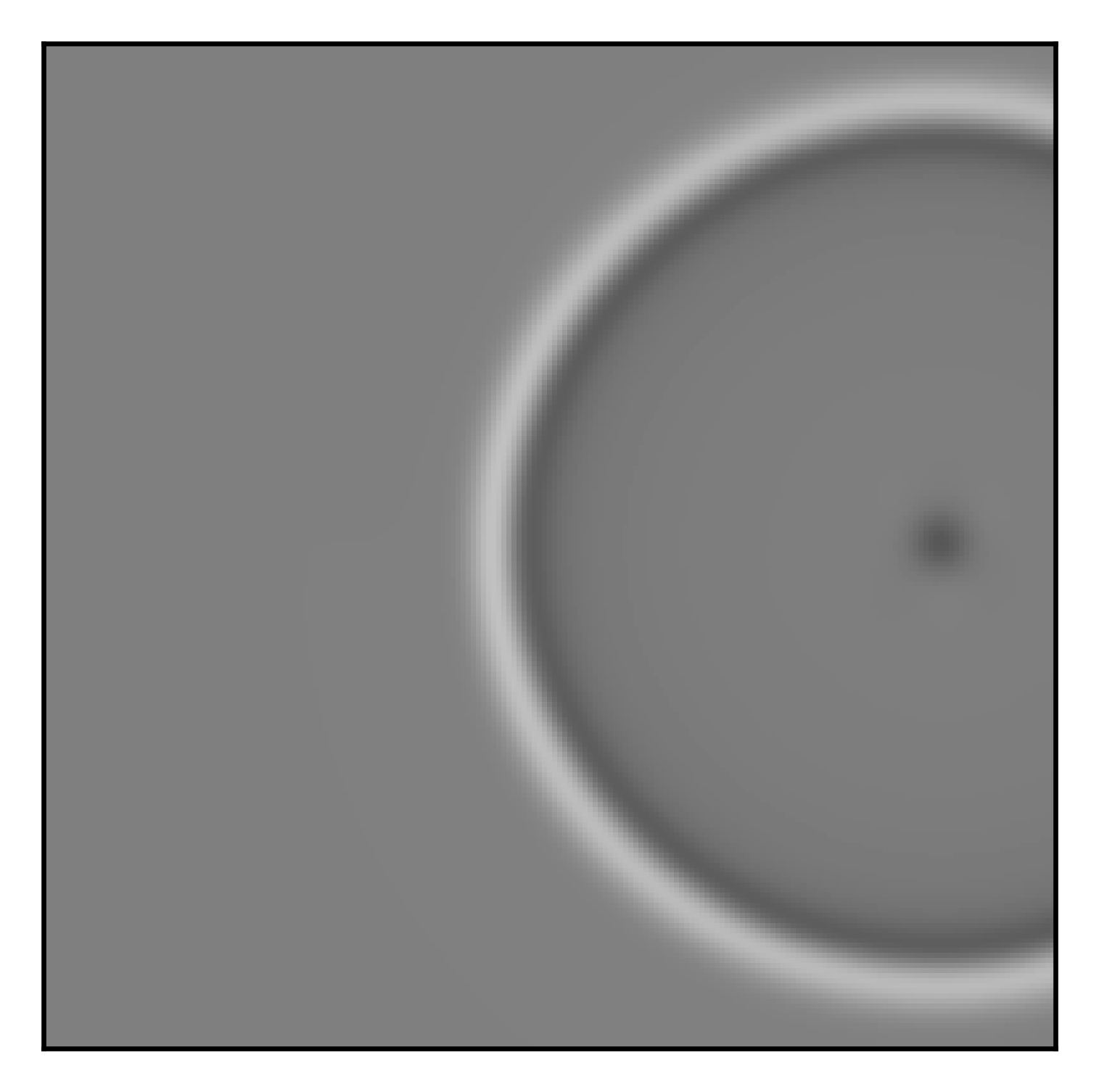}
    \includegraphics[width=\imgwidth]{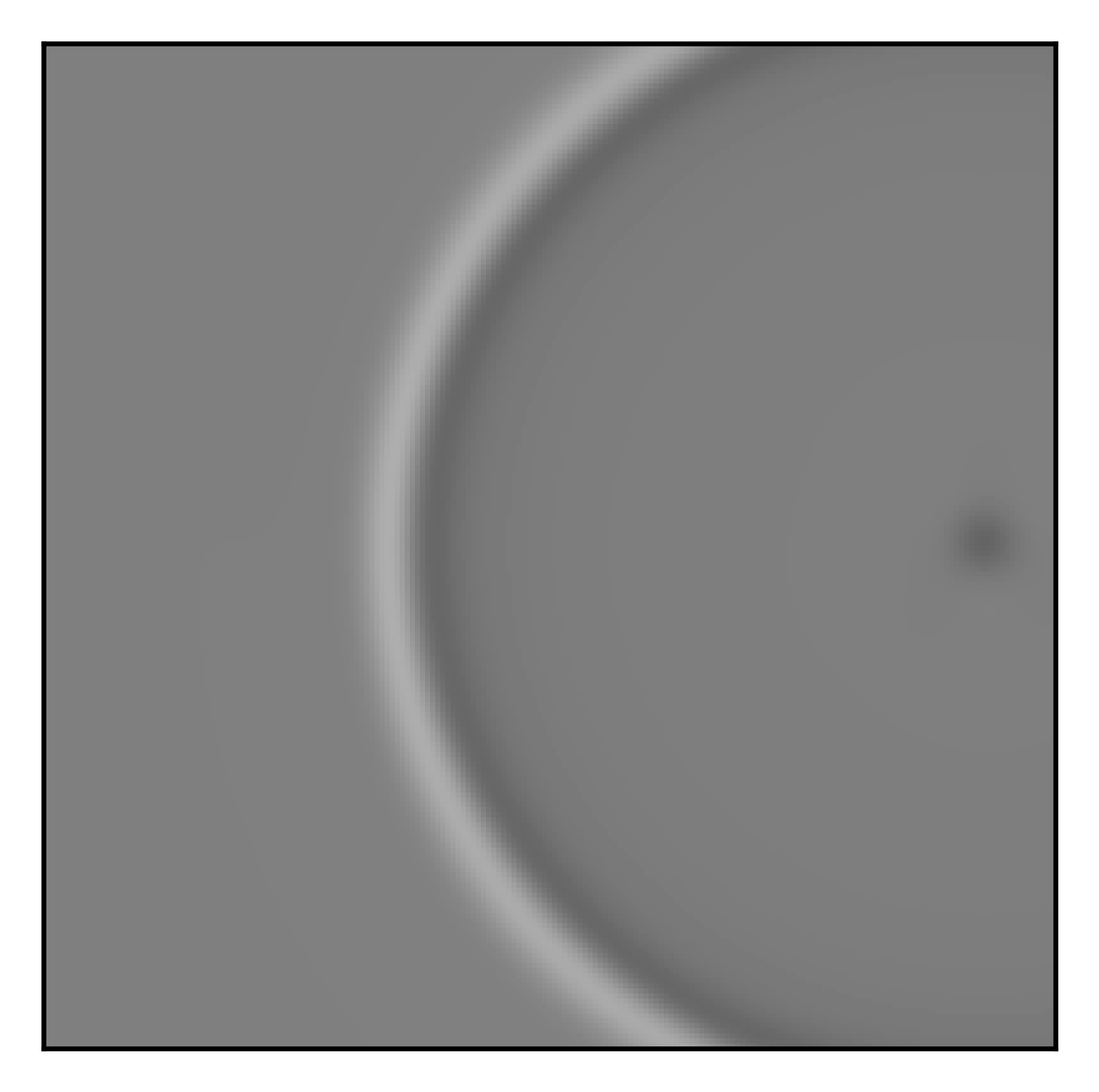}
    \includegraphics[width=\imgwidth]{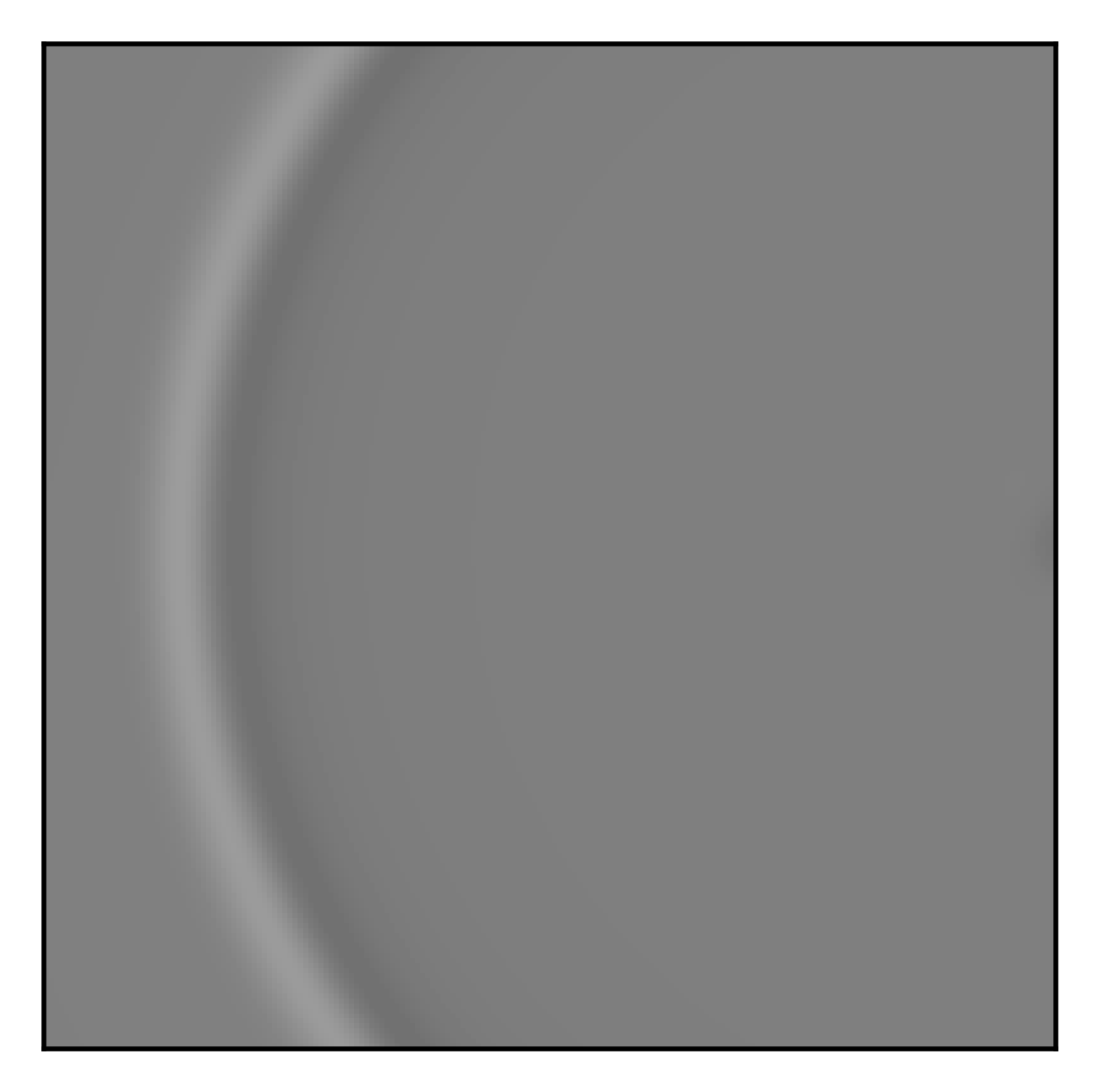}
    \includegraphics[width=\imgwidth]{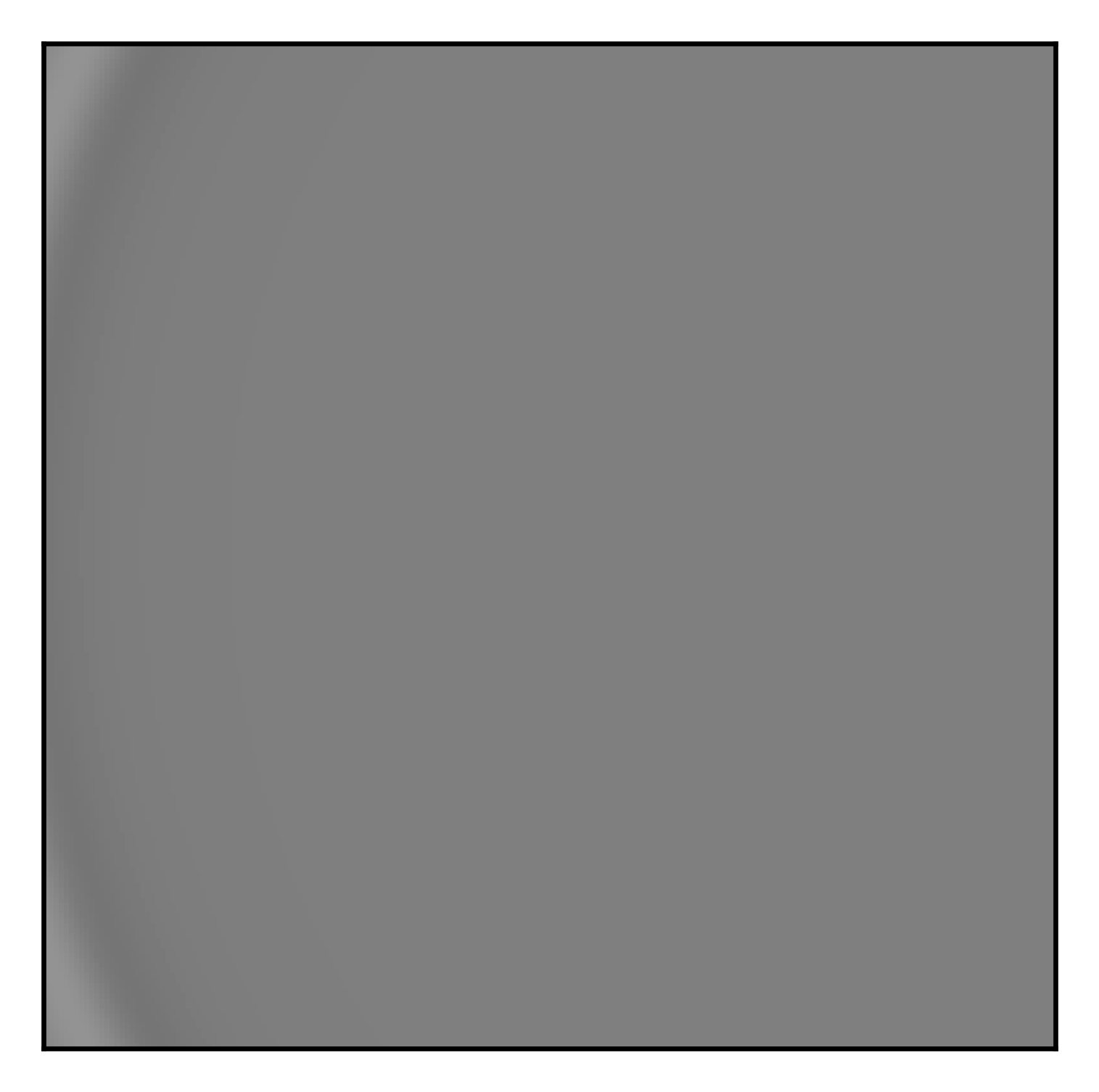}
    \includegraphics[width=\imgwidth]{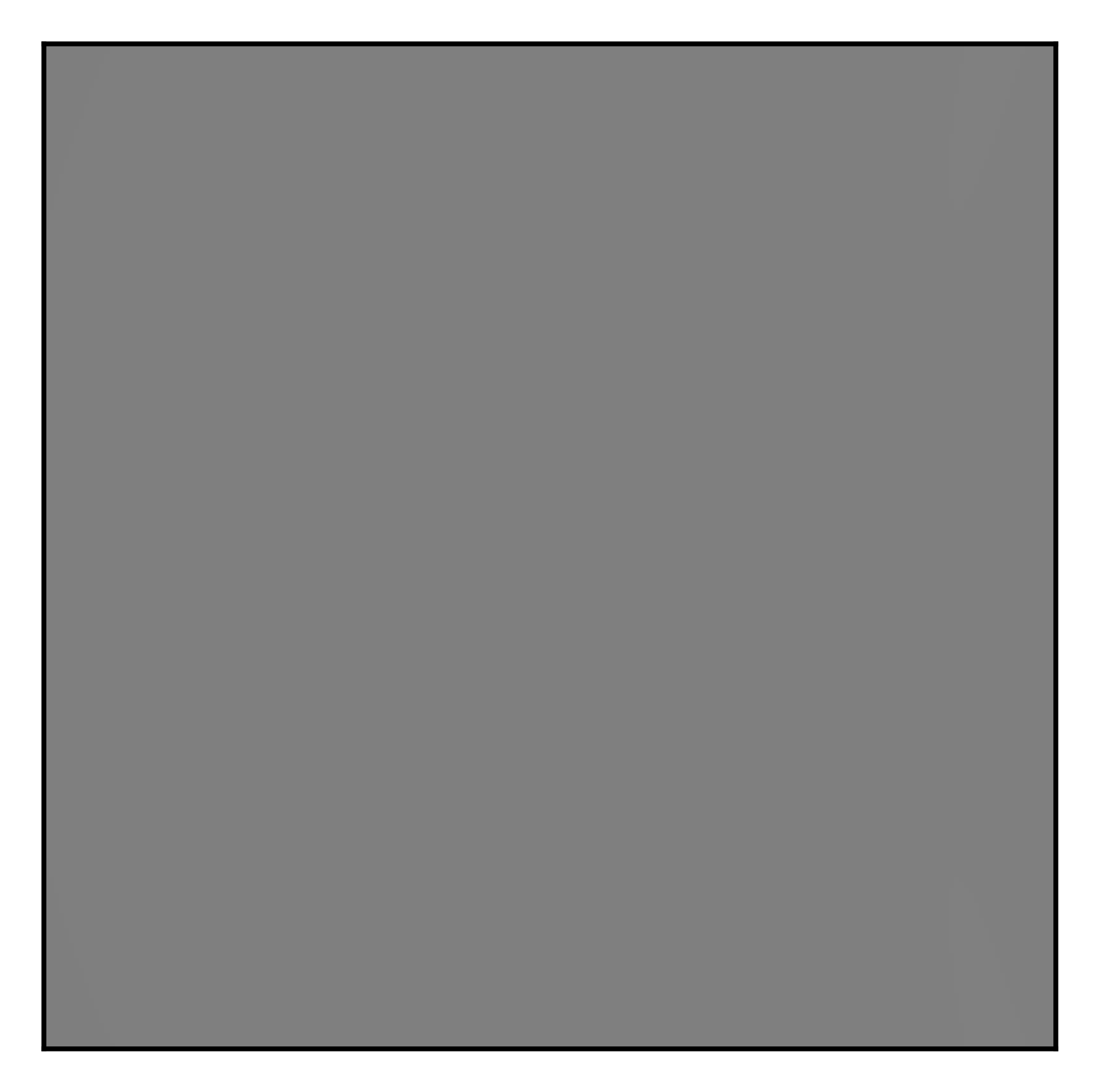}\par\vspace{3pt}

    \parbox{\linewidth}{\centering\textbf{CBC ($\sigma=0, \kappa_2=0$)}}\par\vspace{1pt}
    \includegraphics[width=\imgwidth]{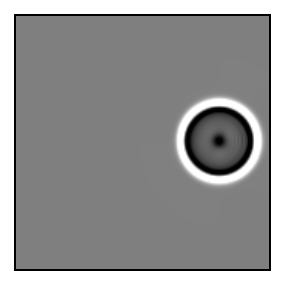}
    \includegraphics[width=\imgwidth]{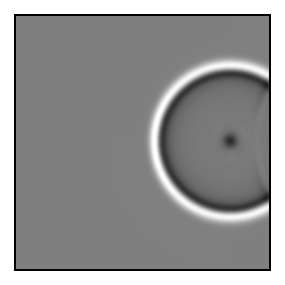}
    \includegraphics[width=\imgwidth]{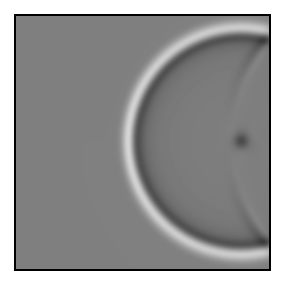}
    \includegraphics[width=\imgwidth]{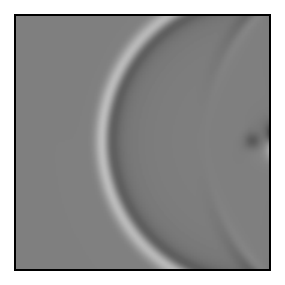}
    \includegraphics[width=\imgwidth]{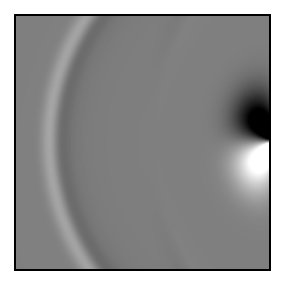}
    \includegraphics[width=\imgwidth]{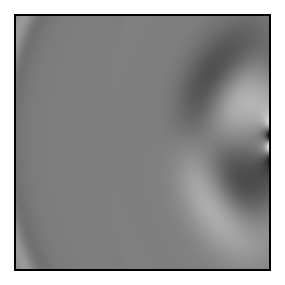}
    \includegraphics[width=\imgwidth]{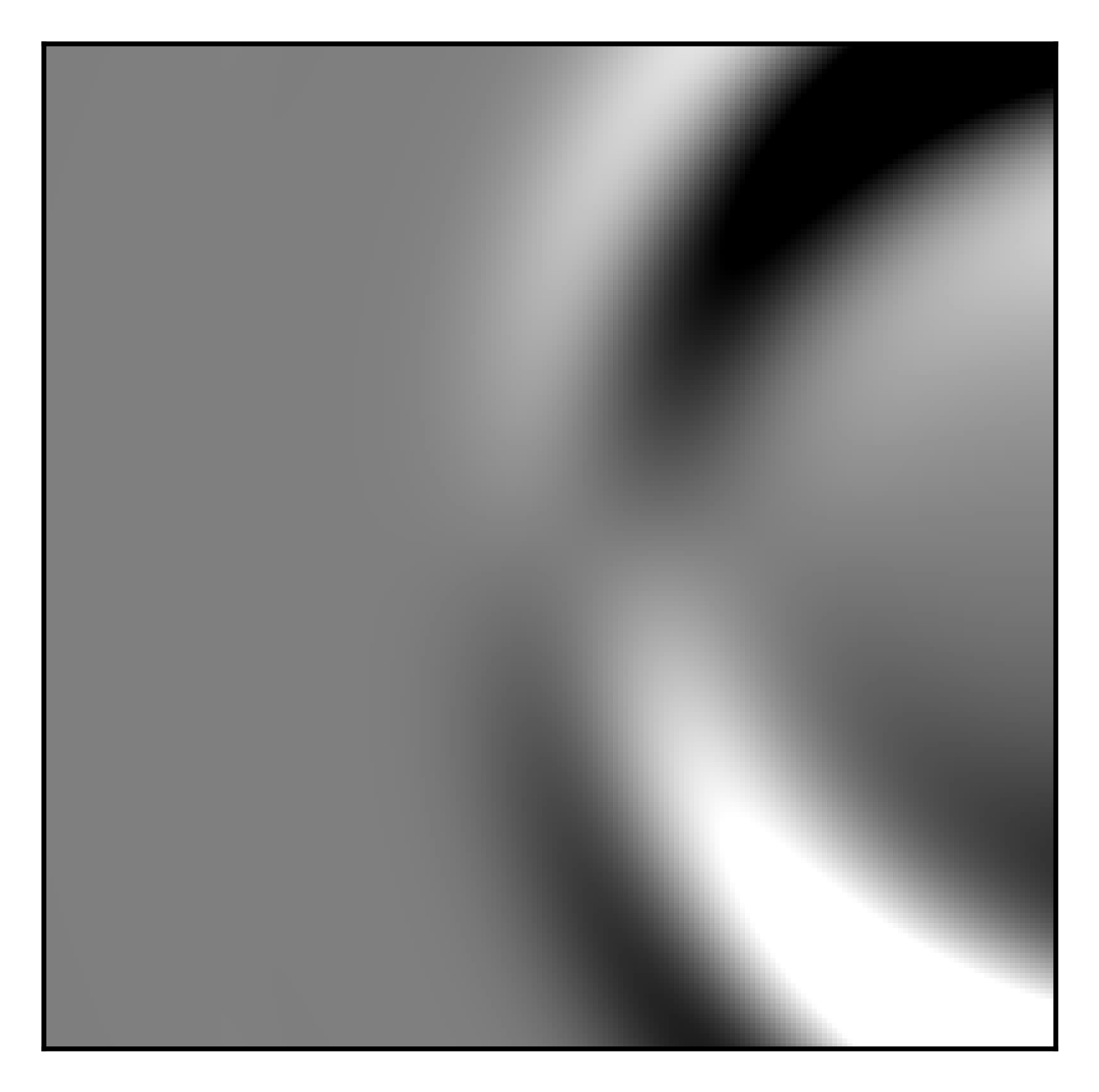}\par\vspace{3pt}

    \parbox{\linewidth}{\centering\textbf{\ac{NN}-CBC}}\par\vspace{1pt}
    \begin{subfigure}[b]{\imgwidth}
        \centering
        \includegraphics[width=\linewidth]{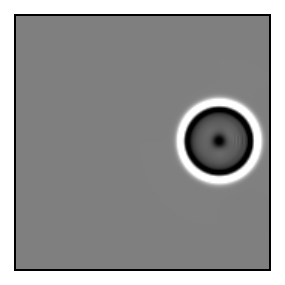}
        \caption*{\scriptsize \hspace{-.3cm} 50} 
    \end{subfigure}
    \begin{subfigure}[b]{\imgwidth}
        \centering
        \includegraphics[width=\linewidth]{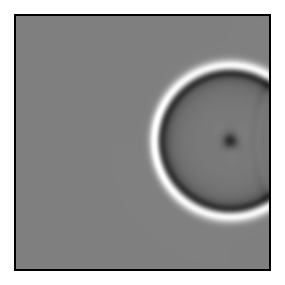}
        \caption*{\scriptsize \hspace{-.3cm} 100 }
    \end{subfigure}
    \begin{subfigure}[b]{\imgwidth}
        \centering
        \includegraphics[width=\linewidth]{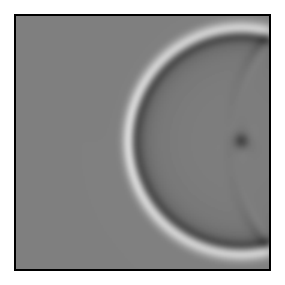}
        \caption*{\scriptsize \hspace{-.3cm} 150}
    \end{subfigure}
    \begin{subfigure}[b]{\imgwidth}
        \centering
        \includegraphics[width=\linewidth]{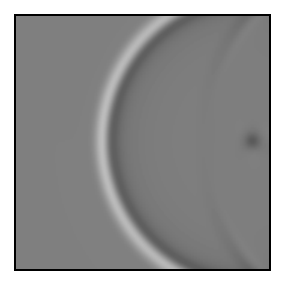}
        \caption*{\scriptsize \hspace{-.3cm} 200}
    \end{subfigure}
    \begin{subfigure}[b]{\imgwidth}
        \centering
        \includegraphics[width=\linewidth]{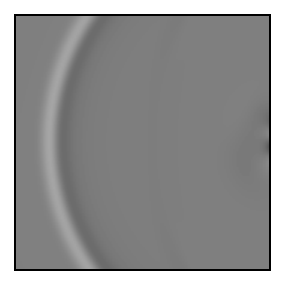}
        \caption*{\scriptsize \hspace{-.3cm} 300}
    \end{subfigure}
    \begin{subfigure}[b]{\imgwidth}
        \centering
        \includegraphics[width=\linewidth]{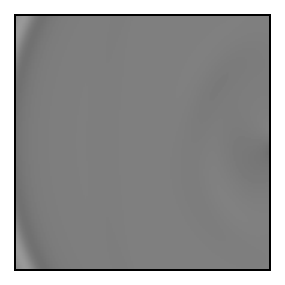}
        \caption*{\scriptsize \hspace{-.3cm} 400}
    \end{subfigure}
    \begin{subfigure}[b]{\imgwidth}
        \centering
        \includegraphics[width=\linewidth]{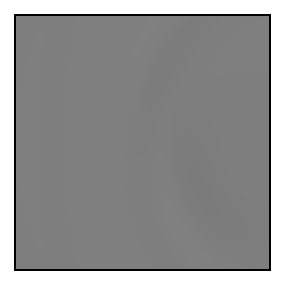}
        \caption*{\scriptsize \hspace{-.3cm} 500}
    \end{subfigure}\par

    \caption{Visual comparison of the density evolution for the convected vortex case. Rows depict different outflow boundary conditions: (Top) Reference simulation with extended domain; (Middle) Characteristic Boundary Condition (CBC) with fixed parameters $\sigma=0, \kappa_2=0$; (Bottom) Neural Network-enhanced \ac{CBC} (\ac{NN}-CBC) with corresponding time steps indicated. The columns represent snapshots in progressive time steps. For a quantitative comparison see Figure \ref{fig:CBC}.}
    \label{fig:PressureWaveEvolution}
\end{figure}

\begin{figure}
    \centering
    \vspace{0.5cm}
    \includegraphics[width=0.49\linewidth]{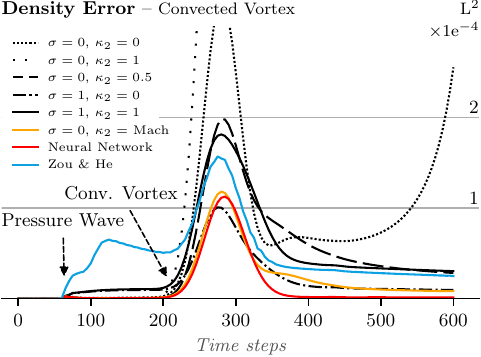}
    \includegraphics[width=0.49\linewidth]{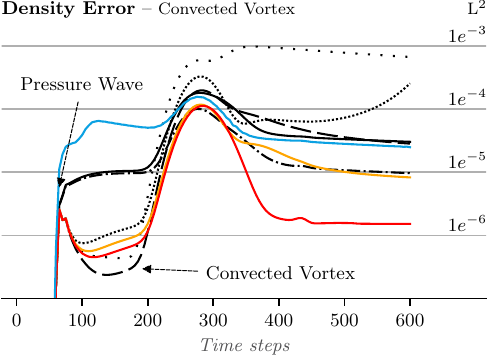}
    \caption{Comparison of the temporal evolution of the L2 norm of the density error between the neural network-enhanced \ac{CBC} and \acp{CBC} of the 2D convected vortex benchmark, depicted for various fixed global parameter sets for $\sigma$ and $\kappa_2$. The error is measured against a reflection-free reference simulation. \textit{Left:}: Linear scaling \textit{Right:}: Logarithmic scaling.}
    \label{fig:CBC} 
\end{figure}

To visually assess the performance of the different boundary conditions, Figure~\ref{fig:PressureWaveEvolution} presents a series of snapshots that illustrate the evolution of the density field for the two-dimensional convected vortex benchmark. This figure compares the behavior of the reference simulation (extended, reflection-free domain), a \ac{CBC} with fixed parameters ($\sigma=0$, $\kappa_2=0$), and the adaptive neural network-enhanced \ac{CBC} (\ac{NN}-CBC) at several representative time steps. Qualitatively, these snapshots reveal differences in how each boundary condition handles the initial pressure wave and, more critically, the main convected vortex as it approaches and leaves the computational domain. The extent of spurious reflections and the integrity of the vortex structure can be visually compared, providing context for the quantitative error analysis. Figure~\ref{fig:CBC} compares the L2 norm of the density error over time for several outflow boundary condition strategies in the 2D convected vortex benchmark. The error is calculated by comparing the simulation results against a reference solution obtained from a simulation with an extended domain, designed to be free of reflections. The figure specifically contrasts simulations using \acp{CBC} with various fixed global parameter sets ($\sigma$ and $\kappa_2$) against the adaptive neural network-enhanced \ac{CBC} and the traditional \textit{Zou \& He} boundary condition.

The results for the \acp{CBC} with fixed parameters, depicted by the black dashed and dotted lines, underscore the inherent challenge of selecting universally optimal global values for $\sigma$ and $\kappa_2$. It is evident that different fixed configurations yield markedly different levels of density error for specific events. This indicates a high sensitivity of the simulation accuracy to these pre-defined settings. For instance, while a $\sigma$ value of zero is often theoretically considered optimal for minimizing reflections, its practical application in this specific \ac{CBC} implementation reveals complexities. In this setup, $\sigma=0$ implies that the \ac{CBC} formulation only involves derivatives, allowing values at the boundary like the pressure to drift. This situation, therefore, does not account for the background state of the flow. Consequently, the lack of coupling can introduce numerical instabilities, particularly when large gradients occur within the characteristic formulation applied at the boundary. This behavior can be observed in Figure~\ref{fig:CBC} for the configurations with $\sigma=0$ and $\kappa\ne 0$. Here, the initial pressure wave caused by the convecting vortex is crossing the boundary at  $t\approx65$ and is seen to experience a lower reflection error than  the other cases. However, these parameter choices subsequently demonstrate poor performance in handling the much stronger convected vortex as it begins to exit the domain around $t\approx200$. The strong deviation from the background flow state leads to a significant increase in density error. In case of missing transversal terms ($\kappa = 0$) in addition, this even results in an increase in density error over time (dotted line). This example illustrates that a parameter value that is beneficial for one type of flow scenario can be detrimental for another.

The capability of the adaptive neural network-enhanced \ac{CBC} (red solid line) is highlighted when the main convected vortex interacts directly with and subsequently passes through the outflow boundary. This is reflected in the consistently low error during this phase ($t>200$). Although the neural networks may not minimize the reflection of the first pressure shockwave as effectively as a specifically tuned  configuration setting $\sigma=0$ directly, it clearly provides a superior performance overall. The constant configuration that best handles the initial wave proves incapable of adequately managing the subsequent more complex vortex. The neural network, in contrast, finds a more robust balance, effectively handling both the initial disturbances and the primary vortical structure. 

Another fixed-parameter configuration worth mentioning is when $\kappa_2$ is set equal to the Mach number, a parametrisation suggested by Wissocq et al.~\cite{Wiscoq2017} (represented by the orange solid line). 
This particular setting demonstrates a commendable ability to handle the initial pressure wave (around $t\approx65$), suggesting its potential as a robust default choice for managing pure wave phenomena. 
Moreover, when compared to the other \ac{CBC} cases with global constant parameters, this approach also performs favorably during the subsequent convected vortex phase ($t\approx 200-300$), exhibiting a lower  error peak than many alternative fixed settings. However, despite these relative advantages over other global \ac{CBC} configurations, its overall performance, particularly for the convected vortex, is still approximately an order of magnitude worse than that achieved by the adaptive neural network-enhanced \ac{CBC} at the end of the simulation.

The traditional \textit{Zou \& He} boundary condition, also included in the comparison, presents its own set of trade-offs. While it demonstrates issues in optimally handling the initial pressure shockwave, exhibiting a notable error peak as this wave interacts with the boundary, its behavior with the main convected vortex offers a different perspective. Although the \textit{Zou \& He} condition leads to significant density errors during the vortex ejection phase, it can be considered a relatively robust or default option when compared to \acp{CBC} that are poorly parameterized. For instance, certain fixed \ac{CBC} settings that might perform well for one specific event (like the initial pressure wave) can lead to substantially larger errors or instabilities when the more complex vortex structure exits the domain. In such scenarios of extreme misconfiguration for a fixed \ac{CBC}, the \textit{Zou \& He} condition, despite its inherent limitations for achieving high accuracy in these sensitive flows, might offer a more predictable, albeit still suboptimal, performance for the main vortex structure.

\acresetall 
\section {Conclusion}
\label{sec:conclusion}
The accurate representation of outflow boundary conditions remains a significant hurdle in computational fluid dynamics. This is particularly evident for the \ac{LBM}, where artificial reflections can compromise the prediction of aerodynamic forces and acoustic emissions. This paper investigated the potential of \acp{NN} to mitigate potential non-physical boundary effects, aiming to enhance simulation accuracy while enabling the use of truncated domains to reduce computational requirements. Two distinct \ac{NN}-based strategies were developed and successfully demonstrated: the direct reconstruction of unknown particle distribution functions at the outflow boundary, and the enhancement of established \acp{CBC} through adaptive parameter tuning.

In the first approach, a direct reconstruction model was trained using data from a two-dimensional flow around a cylinder at a Reynolds number of 200. This \ac{NN}-based boundary condition demonstrated significantly improved predictions for drag, lift, and Strouhal numbers for a range of Reynolds numbers in a truncated domain when compared with the traditional \textit{Zou \& He} condition. Its robustness and generalizability were further demonstrated by its successful application to simulate the flow over a \textit{NACA0012} airfoil. Here, the previously trained boundary condition accurately recovered aerodynamic performance predictions for various angles of attack in a computational domain extending only half a chord length beyond the airfoil. These results underscore the model's ability to learn physically consistent boundary behavior from local flow characteristics and its capability to effectively reduce computational costs without sacrificing physical fidelity. Of particular interest was its ability to accurately suppress non-physical behavior in truncated domains at different Reynolds numbers and with a novel geometry compared to that used in the training data set.

The second strategy focused on augmenting the established \acp{CBC} by employing a neural network to dynamically tune model parameters, $\sigma$ and $\kappa_2$, based on local flow data such as derivatives of pressure and velocity. This approach was evaluated on the benchmark case of a two-dimensional convected vortex. Here, the neural-enhanced \ac{CBC} was able to achieve lower density errors compared to the conventional \acp{CBC} which relies on fixed model parameters. By adapting the boundary parameters in both space and time, the \ac{NN} allows for an optimized response to transient and unsteady flow features, such as pressure waves and vortices that can interact with outflow boundaries and cause non-physical responses.

The findings presented in this work highlight the potential for integrating neural networks with outflow boundary conditions. Both proposed methodologies offer promising pathways to improve the accuracy of aerodynamic and aeroacoustic simulations while concurrently reducing the computational expense associated with typically large domain requirements. Future research should extend these findings to more complex and computationally demanding three-dimensional flow scenarios. Further investigations could also explore enhancing the direct reconstruction of distribution functions by integrating a mechanism that couples the outflow with a predefined far-field atmosphere, for example, by encouraging consistency with a target reference pressure or velocity.

\appendix
\section{Initialize NACA mask} \label{Appendix:Naca0012}
This appendix provides the geometric details of the \textit{NACA0012} airfoil mask used in the simulation setup. The airfoil profile is generated using the standard analytical equations for NACA four-digit airfoils, where the chordwise coordinate typically ranges from $x=0$, the leading edge, to $x=1$, the trailing edge \citep{NASA_NACA0012}.

\begin{equation}
\begin{split}
y = \pm 0.594689181 \left[ \right. 0.298222773\sqrt{x} &-0.127125232 x -0.357907906 x^2 \\
                             & \left. +0.291984971 x^3 -0.105174606 x^4  \right]
\end{split}
\end{equation}

For the \textit{NACA0012}, this formulation defines a symmetrical airfoil with a maximum thickness of $12\%$ of this normalized chord length. The continuous airfoil shape, defined by these normalized coordinates, is then discretized onto the discrete lattice grid used by the \ac{LBM} simulation.  This means the normalized length ($x\in(0,1)$) is mapped and spanned over the specific number of grid points chosen for the airfoil's chord length in the simulation. This process involves identifying lattice nodes that fall within the airfoil boundary and those that represent fluid cells. An interpolation scheme is employed for nodes near the airfoil surface to accurately represent the geometry and establish the fluid-solid interface. The resulting collection of lattice nodes designated as solid forms the bounce-back boundary for the \ac{LBM}, effectively modeling the no-slip condition at the airfoil surface, thus defining the solid-fluid interface for boundary treatment. 

\section*{Acknowledgments}
Travis Mitchell acknowledges the assistance of resources and services from the National Computational Infrastructure (NCI) and the Pawsey Supercomputing Centre in the conception of this research, each with funding from the Australian Government, with the latter also funded by the Government of Western Australia. The authors also acknowledge the early work done by Paulina Tomaszewska who was the student to first test the idea of a neural network reconstructed boundary condition in the TCLB code. Furthermore, simulations for this study were performed using the Platform for Scientific Computing at Bonn-Rhein-Sieg University of Applied Sciences, which is funded by the German Ministry of Education and Research and the Ministry for Culture and Science North Rhine-Westphalia.

\bibliographystyle{elsarticle-num}
\bibliography{bibliography}
\end{document}